\documentclass[journal=jacsat,manuscript=article]{achemso}

\usepackage[dvipsnames]{xcolor} 

\usepackage[version=3]{mhchem} 
\usepackage{amsmath} 
\usepackage{amssymb} 
\usepackage{tabularx} 
\usepackage{booktabs} 
\usepackage{soul}
\usepackage{hyperref}
\usepackage{lineno}

\author{Elton Ogoshi}
\email{elton.ogoshi@ufabc.edu.br}
\author{Henrique Ferreira}
\affiliation{Center for Natural and Human Sciences, Federal University of ABC, Santo André, SP, Brazil}
\alsoaffiliation{Paulista Faculty of Informatics and Administration, São Paulo, SP, Brazil}
\author{João N. B. Rodrigues}
\affiliation{Center for Natural and Human Sciences, Federal University of ABC, Santo André, SP, Brazil}
\alsoaffiliation{Physics Departament, Federal University of Pernambuco, Recife, PE, Brazil}
\author{Gustavo M. Dalpian}
\email{dalpian@if.usp.br}
\affiliation{Center for Natural and Human Sciences, Federal University of ABC, Santo André, SP, Brazil}
\alsoaffiliation{Institute of Physics, University of São Paulo, São Paulo, SP, Brazil}

\title{Exploring chemical compound space with a graph-based recommender system}


\begin{document}






\begin{abstract}
  
With the availability of extensive databases of inorganic materials, data-driven approaches leveraging machine learning have gained prominence in materials science research. In this study, we propose an innovative adaptation of data-driven concepts to the mapping and exploration of chemical compound space. Recommender systems, widely utilized for suggesting items to users, employ techniques such as collaborative filtering, which rely on bipartite graphs composed of users, items, and their interactions. Building upon the Open Quantum Materials Database (OQMD), we constructed a bipartite graph where elements from the periodic table and sites within crystal structures are treated as separate entities. The relationships between them, defined by the presence of ions at specific sites and weighted according to the thermodynamic stability of the respective compounds, allowed us to generate an embedding space that contains vector representations for each ion and each site. Through the correlation of ion-site occupancy with their respective distances within the embedding space, we explored new ion-site occupancies, facilitating the discovery of novel stable compounds. Moreover, the graph's embedding space enabled a comprehensive examination of chemical similarities among elements, and a detailed analysis of local geometries of sites. To demonstrate the effectiveness and robustness of our method, we conducted a historical evaluation using different versions of the OQMD and recommended new compounds with Kagome lattices, showcasing the applicability of our approach to practical materials design.

\end{abstract}

\section{Introduction}

The set of all admissible compounds results from all possible combinations of atoms, compositions\cite{all_materials2016} and crystal structures\cite{oganov2006crystal} (ACS\cite{Zunger2018}). The combinatorially large nature of this space makes it extremely difficult to thoroughly explore it. A naive approach to expand the quantity of available known materials would be a trial-and-error methodology across all possibilities, but this is is rendered impractical due to resource limitations, a constraint that holds true even for \textit{in silico} exploration. Despite the significant recent advancements in high-throughput infrastructure and computational power, the most effective strategy for researchers remains the identification of patterns within known compounds to effectively restrict this expansive search space.

In the past, there were some attempts to formalize such patterns in the form of empirical geometrical rules, such as the Goldschmidt tolerance factor~\cite{goldschmidt1926} for perovskites, and the Hume-Rothery rules for the formation of substitutional solid solutions in metallic systems \cite{callister_book}. However, when these empirical rules were formulated, the available dataset of materials  was considerably limited compared to the vast repositories we have today. The advent of comprehensive materials databases, such as the Open Quantum Materials Database (OQMD)~\cite{Saal2013}, Aflowlib~\cite{Curtarolo2012}, Materials Project~\cite{Jain2013}, and NOMAD\cite{Draxl2019} has significantly expanded the compounds' knowledge base. This broadens the prospects for both finding new empirical rules and improving the existing ones. For instance, a larger dataset of compounds allowed Bartel et al~\cite{Bartel2019} to propose a new and more accurate tolerance factor for perovskites, while  Pei et al~\cite{ml_alloy2020} found new descriptors that predicted the stability of high-entropy alloys beyond the Hume-Rothery rules. Yet, these approaches are typically limited to a specific subset of compounds constrained by a crystal structure.

In recent years, inspired by data-driven methodologies, researchers have proposed innovative strategies to narrow the search space across a broader set of compounds. These strategies can be broadly categorized into iterative and non-iterative methods. Iterative methods employ a cyclical process of exploration and exploitation, with examples including active learning~\cite{montoya2020autonomous}, genetic algorithms~\cite{uspex2013}, and iterative ion substitution~\cite{ion_substitution2021}. Non-iterative methods, on the other hand, build models from the entirety of the available data. Notable such examples include the tensor-based recommender system proposed by Seko et al~\cite{tensor_recommender2018, hayashi2022recommender}, and state-of-art graph-based machine learning models that predict the formation energy of unexplored compounds, thereby helping to restrict the search space~\cite{graph_review2022, m3gnet2022, alignnff_2023}.

Another context where one has to cope with an enormous informational space is the realm of digital life. There users are routinely presented with a vast array of options, be it media to consume or products to purchase. They have to parse through all that information and make decisions that are most aligned with their interests. Some of the most popular information filtering strategies are known as recommender systems~\cite{recommendation_survey2022} (RS). These attempt to simplify the decision-making process by first trying to identify items that are perceived to most closely align with the user's interests and then suggesting these to the user.

Collaborative filtering, a method used by RSs~\cite{collab2009}, makes predictions about the interests of a user by collecting preferences from many users. This process can be visualized and modeled using a bipartite graph, where one set of nodes represents the users, and the other set represents the items. The task of the RS then becomes a link prediction problem: based on the existing structure of the graph and the preferences of similar users, which new user-item links are most likely to form?

In this study, inspired by information filtering ideas, we propose a new way of looking at chemical compound space and apply recommender system (RS) principles to the search for new materials. We construct a bipartite graph to depict the relations between periodic table elements and local crystal environment geometries (ignorant of chemistry). By then generating an embedding space from this graph, we are able to explore the affinities between ions and local environments, and eventually spot new stable materials

In the remainer of this text we will use the expression \textit{anonymous motif sites} (AM sites) to refer to the local crystal environment geometries mentioned in the previous paragraph. We discuss AM sites in detail in the \hyperref[amsites]{Methods} section. Within our model, a ion-AMsite link means that that ion occupies that AM site in one or more materials of the set of compounds used to construct the graph. The weight assigned to each link corresponds to the respective compound's thermodynamic stability, as derived from Density Functional Theory (DFT). By embedding this representation into a vector space, we were able to discern chemical and structural patterns and develop a RS that recommends new ion-site occupations. These recommendations  may result in the formation of stable compounds within a desired prototype crystal structure, as we demonstrate in section \hyperref[kagomeprediction]{Results and Discussion} for the case of \ce{CsV3Sb5}\cite{csv3sb5_2020, kagome2019}.

Our methodology notably differs from previous machine learning and data-driven methods that have generated element embeddings from the periodic table\cite{jha2018elemnet, zhou2018learning, schutt2018schnet, antunes2022distributed}. Unlike those approaches, our embedding for ions is constructed exclusively from the data of ion-site occupancies in known compounds, without considering any compositional descriptor or chemical features from the elements themselves. This unique approach emphasizes the incorporation of site information within the embedding and, as a consequence, we  cannot only delve into ion-ion relationships within the embedding, but also exploit and analyse the ion-site relationships.

\section{Methods}

\subsection{The Link Prediction Task in Bipartite Graphs}

Link prediction can be formally described as follows: consider a graph $G = (V, E^+)$, where $V$ is the set of nodes and $E^+$ the set of edges. In the context of a recommender system, $G$ is a bipartite graph with $V = U \cup P$ and $U \cap P = \emptyset$. Here, $U$ represents the set of users and $P$ the set of products, for example. Each edge connects a vertex in $U$ to a vertex in $P$, with edges $(u, p) \in E^+$ representing user-product interactions.

\begin{sloppypar}
The set of all possible but non-existing edges is denoted by $E^-$, such that ${E^+ \cap E^- = \emptyset}$. The link prediction task can then be formulated as a binary classification problem, where all allowed edges or pairs of nodes $(u, p) \in U \times P$ are associated with a label $y_{up} \in \{0, 1\}$, indicating the absence or presence of an edge between $u$ and $p$, respectively. 
\end{sloppypar}

The objective of the prediction task is to learn a function $f: U \times P \rightarrow [0, 1]$ that approximates the probability $P(y_{up} = 1 \mid u, p)$ of an edge existing between nodes $u$ and $p$.

\begin{equation}
P(y_{up} = 1 | u, p) \approx f(u, p)
\end{equation}

\begin{figure*}[ht]
    \centering
    \includegraphics[width=14cm]{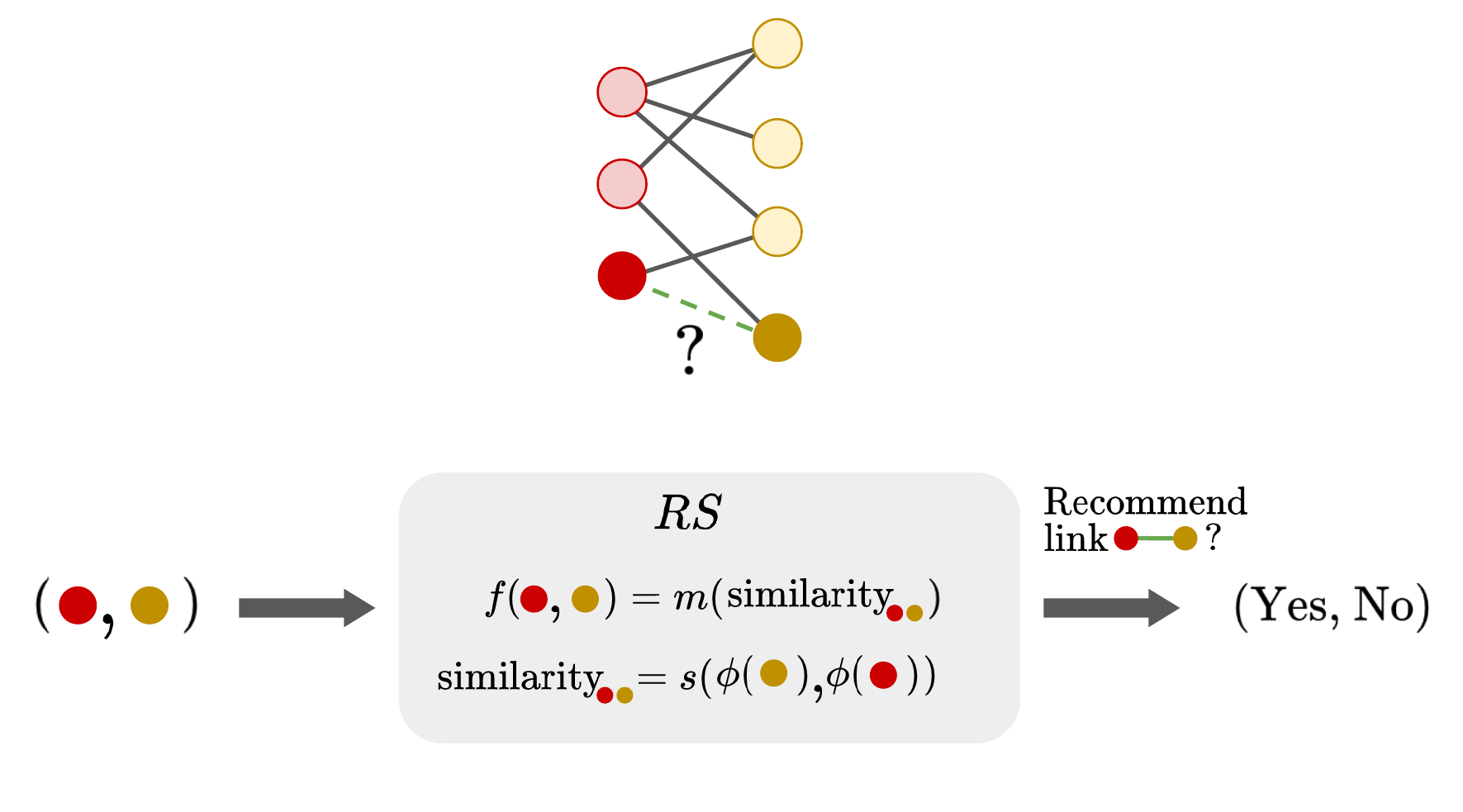}
    \caption{Visualization of the link prediction task for recommender systems ($RS$), encapsulated by function $f$. The goal is to suggest new candidate connections between a pair of nodes, represented in deep red and deep yellow. The function $f$ is composed of two components. Firstly, the nodes are processed by an embedding function $\phi$ which converts each node into a vector representation. Then, these vectors are used as input for a similarity function $s$. The output from the similarity function is then processed by a model $m$, which finally produces the recommendation. Through this process, the recommender system is capable of suggesting new links between nodes.}
    \label{fig:link_prediction}
\end{figure*}

A common approach for link prediction employs node embeddings, that are vector representations of nodes that capture their local graph structural information. Let $\phi: V \rightarrow \mathbb{R}^d$ be a function mapping each node $v \in V$ to a $d$-dimensional vector $\phi(v)$. This function outputs vector representations for both $U$ and $P$ ($\boldsymbol{v}_u$ and $\boldsymbol{v}_p$, respectively). The similarity between the embeddings of two distinct nodes can be computed using a similarity function $s: \mathbb{R}^d \times \mathbb{R}^d \rightarrow \mathbb{R}$, such as the dot product or the cosine similarity. This can then be used as a proxy for the likelihood of a link existing between them, or used as an input to a second trained model $m$, such as a decision tree or a logistic regression, responsible for defining the similarity threshold that delineates the existing edges from the non-existing edges. The sequence of transformations from the node representations to the final link prediction function $f$ is represented in Figure \ref{fig:link_prediction}.

To train a link prediction model $f$, a training dataset $\mathcal{D} = {\{(u_i, p_i, y_{u_i p_i})\}}_{i=1}^{N}$ of $N$ instances is constructed from positive examples $(u, p, y_{up} = 1)$ drawn from the true edges $E^+$ and negative examples $(u, v, y_{up} = 0)$ drawn from the non-existing edges $E^-$. The model parameters are optimized by minimizing a global loss function $L$ that measures the discrepancy between the predicted probabilities $f(u, p)$ and the true labels $y_{up}$. The edges from the set of positive examples are removed from $G$ to avoid data leakage during the optimization and training of $m$. The dataset $\mathcal{D}$ is partitioned into training, validation, and test sets, ensuring a balanced distribution of positive and negative samples. The global loss function can be written as

\begin{equation}
L = \sum_{(u, p) \in \mathcal{D}} \mathcal{L}(f(u, p), y_{up})
\end{equation}

where $f(u, p)$ is the predicted probability of edge existence, $y_{up}$ is the true label, and $\mathcal{L}$ is the loss function defined by the model $m$.

The learned function $f$ is evaluated on the test set to estimate its generalization ability to unseen data. Ultimately, $f$ is used to estimate the probability of all edges $(u, p) \in E^-$ for new recommendations.

\subsection{Building the Recommender System}
\subsubsection{The Data:} We utilized two versions of the OQMD database: version 1.4 from 2020 and version 1.5 from 2021, denoted as $\text{OQMD}^{\text{past}}$ and $\text{OQMD}^{\text{current}}$, respectively. Each version was subjected to the following filters:

\begin{enumerate}
\item Non-unary materials with negative formation energy ($E_f < 0.0\ \text{eV}$) were selected. For unary materials, no restrictions were applied as the energy above hull $E_{\text{hull}}$ is equivalent to $E_f$.
\item Only unary, binary, ternary, and quaternary structures with 24 or fewer sites in the unit cell were considered. This step was applied to try to eliminate structures with a high number of low-symmetry non-equivalent occupied sites, such as in amorphous phases.
\item Duplicate entries were removed, i.e., data entries reported by OQMD as repeated for the same compound (indicated by the column \textit{duplicate\_of\_id}).
\end{enumerate}

After applying these filters, $\text{OQMD}^{\text{past}}$ contained 307,912 entries and $\text{OQMD}^{\text{current}}$ contained 451,522 entries. The energy above hull $E_{\text{hull}}$ and the crystal structure were extracted for each compound in both databases.

Our first objective is to build a recommender system using $\text{OQMD}^{\text{past}}$ and then validate it by verifying two different things: (i) how much do the recommendations for ion-site occupations restrict the search space; (ii) how many of the thermodynamically stable coumpounds from $(\text{OQMD}^{\text{current}}-\text{OQMD}^{\text{past}})$ can be found in this restricted search space. For simplicity, we carried out the latter validation experiment on a subset of cubic halide perovskites, rather than using the entire set of compounds $\mathcal{C}$. This process, often referred to as "historical validation", also helped identify the optimal set of hyperparameters for the model and the graph $G$ construction.

Following this validation step, we built a recommender system using $\text{OQMD}^{\text{current}}$ data and generated recommendations for ion-site occupations. These recommendations resulted in the proposal of new compounds in a crystal structure prototype. Subsequently, we conducted Density Functional Theory (DFT) calculations, using the same parameters of OQMD, and utilized OQMD's convex hull constructions to calculate the recommended compounds' $E_{\text{hull}}$. This allowed us to infer about the thermodynamic stability of the new compounds and from it validate the RS recommendations.

\subsubsection{Anonymous Motifs (AMs)}
\label{am}

Here we present the definition of an Anonymous Motif (AM), since to the best of our knowledge, a formal mathematical definition for AMs does not exist in the literature of crystallography. 

AM is defined solely by the geometric arrangement and connectivity of atomic positions in space, disregarding the chemical identity of the atoms and, consequently, the compound's stoichiometry. Given a compound’s crystal structure, we first disregard its composition and reduce it, when possible, to a primitive cell $p$. Then, we use spglib\cite{spglib} to analyze the symmetry of the resulting structure and of its sites. With the symmetry analysis data, we then define AM as the following tuple:

\begin{equation}
AM = (sg, n_s, Wy, l)
\end{equation}

where:
\begin{itemize}
\item $sg$ is the space group number (integer) of $p$
\item $n_s$ is the number of sites in $p$ (integer)
\item $Wy$ is a list of non-equivalent sites in $p$, given by their respective Wyckoff positions, represented as tuples of the form $(n_i, sym_i)$, where $n_i$ is the occupancy count of site $i$ (integer) and $sym_i$ is the Hermann-Mauguin symbol for the point group symmetry of site $i$ (string) in $p$.
\item $l$ is a subgroup number (integer), assigned after using pymatgen's StructureMatcher\cite{pymatgen} to further differentiate the $p$'s with the same $(sg, N, Wy)$ but with different coordination environments. This is a "dummy index", as it is used purely to differentiate otherwise identical AM labels that represent different coordination environments. 
\end{itemize}

\begin{figure*}[ht]
    \centering
    \includegraphics[width=\textwidth]{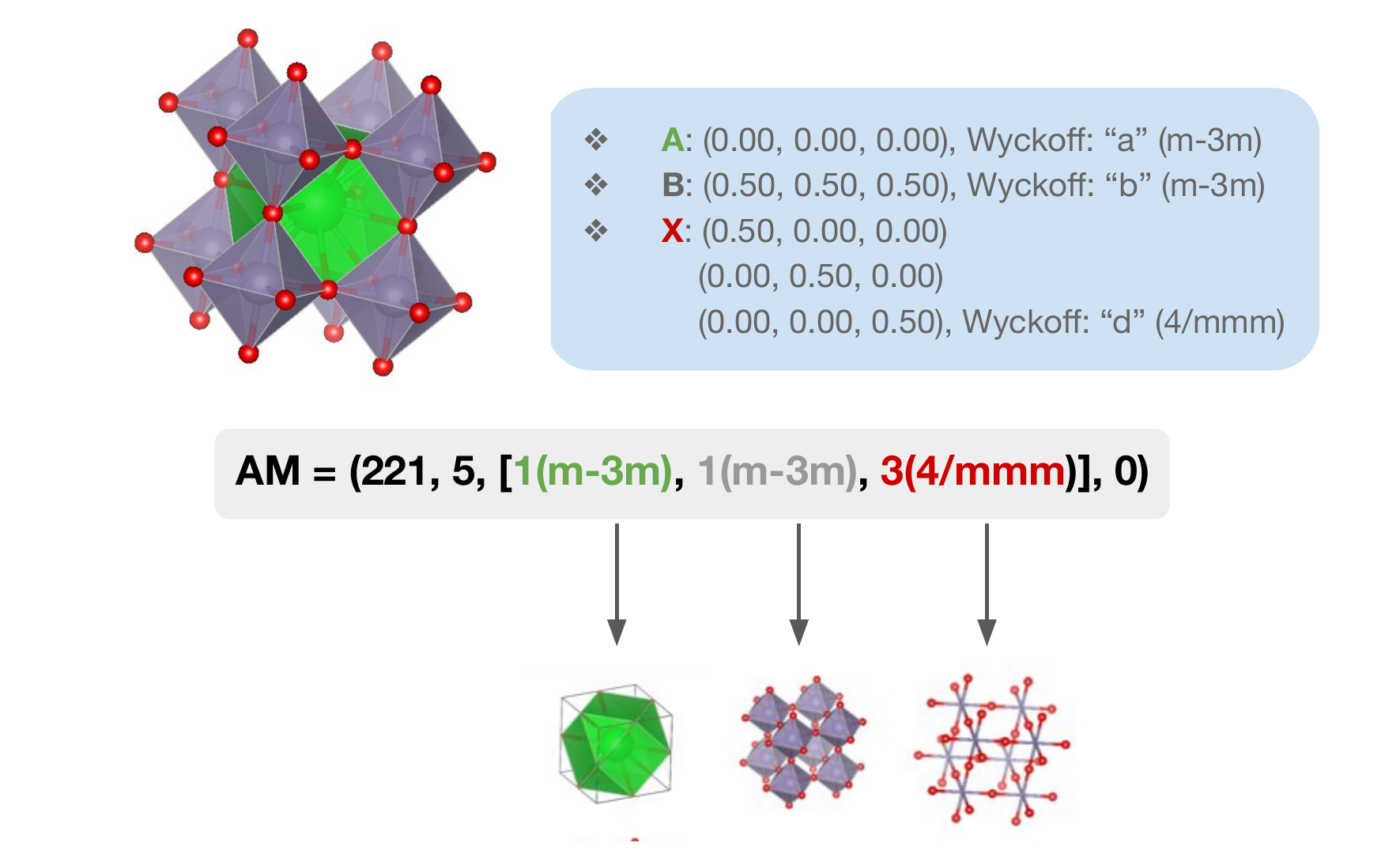}
    \caption{Illustrative example of the Anonymous Motif label of cubic perovskites. The elements of the $Wy$ list represents the A, B, and X sites.}
    \label{fig:perovskite_AM}
\end{figure*}

\begin{sloppypar}
As an illustrative example, we can consider the representation for cubic perovskites: ${(221, 5, [(1, \text{m-3m}), (1, \text{m-3m}), (3, 4/\text{mmm})], 0)}$. The three labels in $Wy$  represent the cuboctahedral A-site, the octahedral B-site, and the linear X-site, respectively (as shown in Figure~\ref{fig:perovskite_AM}. As the chemical identities are disregarded, both simple cubic perovskites and double perovskites would be grouped together into the same AM. By using AMs, one can reduce the complexity of the materials space by identifying common coordination patterns and geometrical motifs shared among different compounds. They provide a standardized way of categorizing materials based on their crystal structures and occupied sites.
\end{sloppypar}

\subsubsection{AM sites}
\label{amsites}

We define $\mathcal{C}$ as the set of all unique compounds $c$ from the OQMD data. Let's denote the set of all AMs found for all compounds in $\mathcal{C}$ as $AM^\mathcal{C}$. Now, we can define the set $S^\mathcal{C}$ of all sites of all AMs in $AM^\mathcal{C}$.

\begin{equation}
S^\mathcal{C} = \{(sg, n_s, n_i, sym_i, l) \mid \forall AM \in AM^\mathcal{C}, (n_i, sym_i) \in Wy\}
\end{equation}

\begin{figure*}[ht]
    \centering
    \includegraphics[width=16cm]{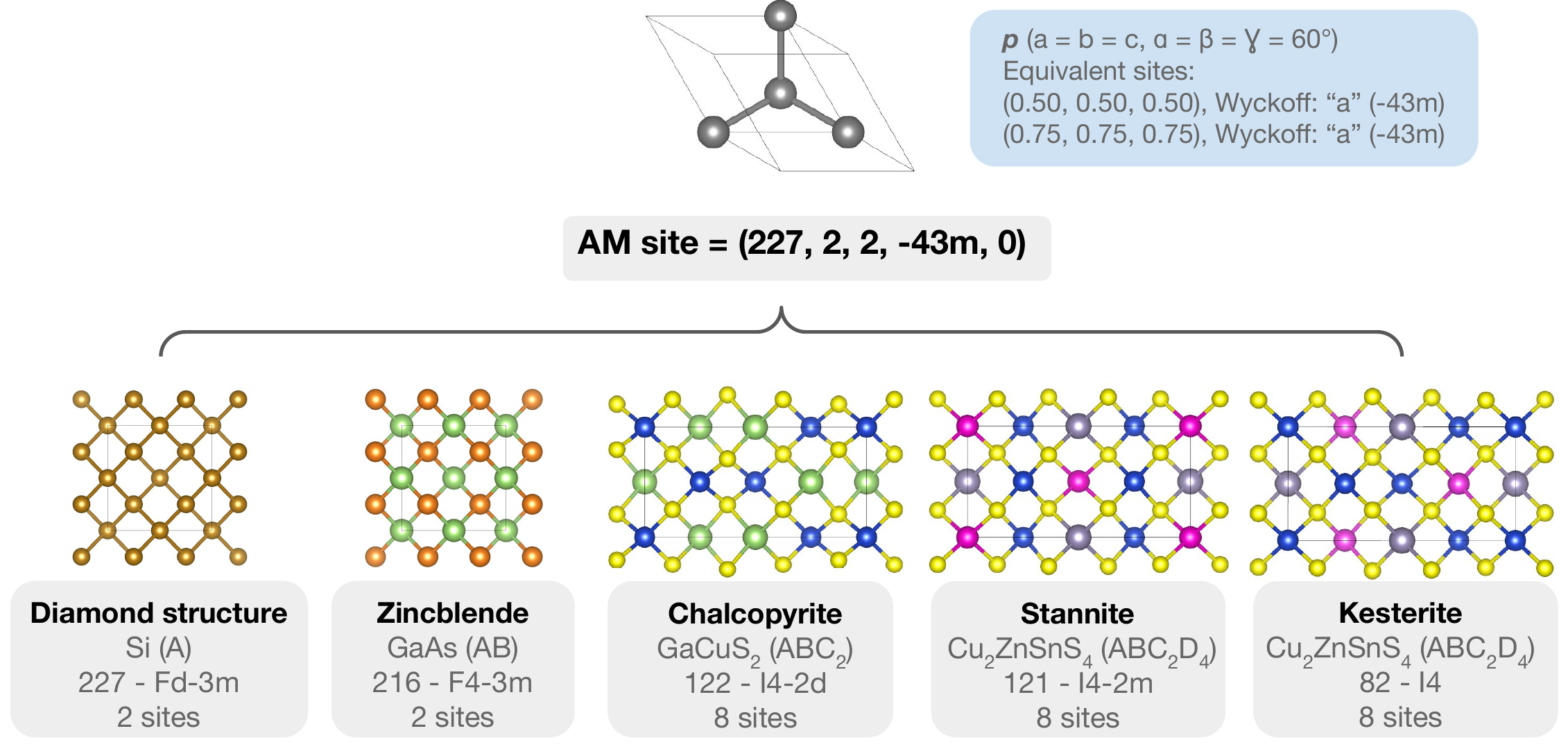}
    \caption{AM site grouping example. As only the lattice geometry is taken into account by ignoring compositions and stoichiometries, the occupied sites of all the presented compounds get mapped to the same (227, 2, 2, -43m, 0) AM site.}
    \label{fig:anonymous_motif_site}
\end{figure*}

As an example, the AM site $(227, 2, 2, -43m, 0)$ groups together all sites of silicon, zincblende, chalcopyrite, stannite and kesterite structures, as shown in detail in Figure~\ref{fig:anonymous_motif_site}. For statistical reasons, we only considered AMs with more than 5 compounds.

\subsubsection{Defining the Bipartite Graph $G^{T}$}

We denote the set of all ions in the chemical compositions of all compounds in $\mathcal{C}$ as $I$. The bipartite graph is constructed with vertices divided into two disjoint sets, $I$ and $S^\mathcal{C}$, where each edge connects a vertex in $I$ to one in $S^\mathcal{C}$. The graph is represented as $G^T=(I, S^\mathcal{C}, E^+, W^T)$, where $E^+$ is the set of existing edges, and $W^T: E^+ \rightarrow \mathbb{R}$ is a Boltzmann-like weight function defining the edge weights. 

The link $(i,s)$ is added to $E^+$ if the occupation of an AM site $s$ by an ion $i$ occurs in a compound $c$. The set of all links is thus defined as:

\begin{equation}
E^+ = \{(i, s) \mid i \text{ occupies } s \text{ in compound } c,  \forall c \in \mathcal{C}\}
\end{equation}

A weight $w^{c,T}(i, s)$ is defined as being proportional to the energy above the hull of the compound $c$, $E_{\text{hull}}^c$, using a Boltzmann factor and introducing a parametric temperature $T$:

\begin{equation}
    w^{c,T}(i, s) = \exp(-E_{\text{hull}}^c / kT)
\end{equation}

\begin{figure*}[ht]
    \centering
    \includegraphics[width=\textwidth]{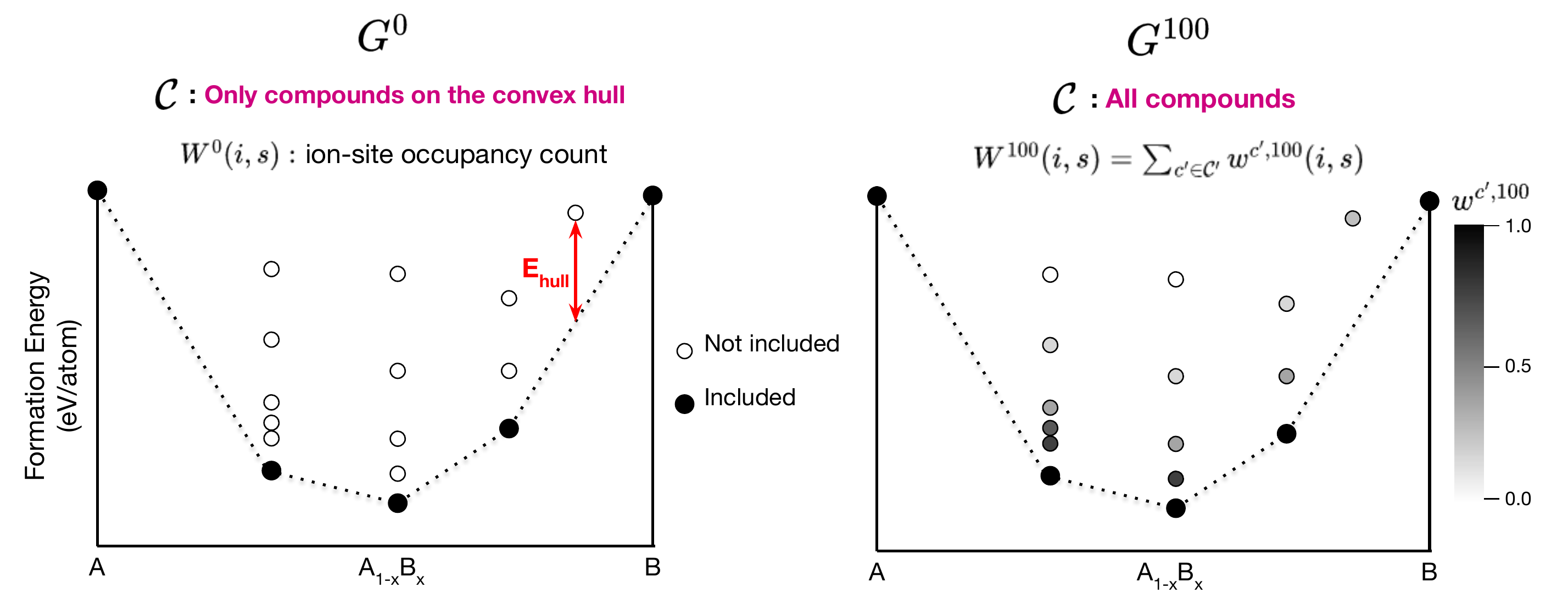}
    \caption{Illustration of the effect of the temperature $T$ on the selection of the compounds $\mathcal{C}$ used for building $G^T$, and on the weights $W^T(i,s)$ between the ion $i$ and the site $s$.} 
    \label{fig:graph_weights}
\end{figure*}

We created three graphs $G^{T}$ with three parameteric temperatures: $0\ \text{K}$, $100\ \text{K}$, and $300\ \text{K}$. The objective was to define a threshold for compounds $c \in \mathcal{C}$: a 0K parametric temperature includes only compounds in the convex hull (here defined as the compounds with $E_{\text{hull}} \leq 30 \text{ meV/atom}$), while  T $>$ 0 K parametric temperatures attribute a small weight even for compounds with large $E_{\text{hull}}$. We wanted to investigate if adding these compounds into the model would improve the recommendations. Larger temperatures penalizes less the weights of the graph links associated with compounds that are thermodynamically unstable. Figure~\ref{fig:graph_weights} illustrates this concept on a hypothetical binary convex hull.

Each edge's total weight $W^T(i, s)$ is computed iteratively over all compounds $c'$ of the set of compounds $\mathcal{C}'$, where $\mathcal{C}'$ is the subset of all compounds that are formed by the occupation of $s$ by $i$:

\begin{equation}
W^T(i, s) = \sum_{c' \in \mathcal{C}'} w^{c',T}(i, s)
\end{equation}

We truncated the values of $w^{c',T}$: if $w^{c',T}(i, s) < 0.001$, then we considered that $(i, s)$ is non-existent.

\begin{sloppypar}
This workflow assigns larger weights $W^T$ to common occupations, such as the X-site in cubic perovskites ${(\text{O}, (221, 5, 3, (3, 4/\text{mmm}), 0))}$ that occur in numerous oxide cubic perovskites, per comparison with the less common occupations like ${(\text{Ca}, (221, 5, 3, (3, 4/\text{mmm}), 0))}$, found in the anti-perovskite $\text{Ca}_3\text{SnO}$. The weight $W^T(i,s)$ quantifies the frequency of the $(i, s)$ occupation in $\mathcal{C}$, weighted by the thermodynamic stability of the compounds featuring $(i, s)$.
\end{sloppypar}

\begin{figure*}[ht]
    \centering
    \includegraphics[width=\textwidth]{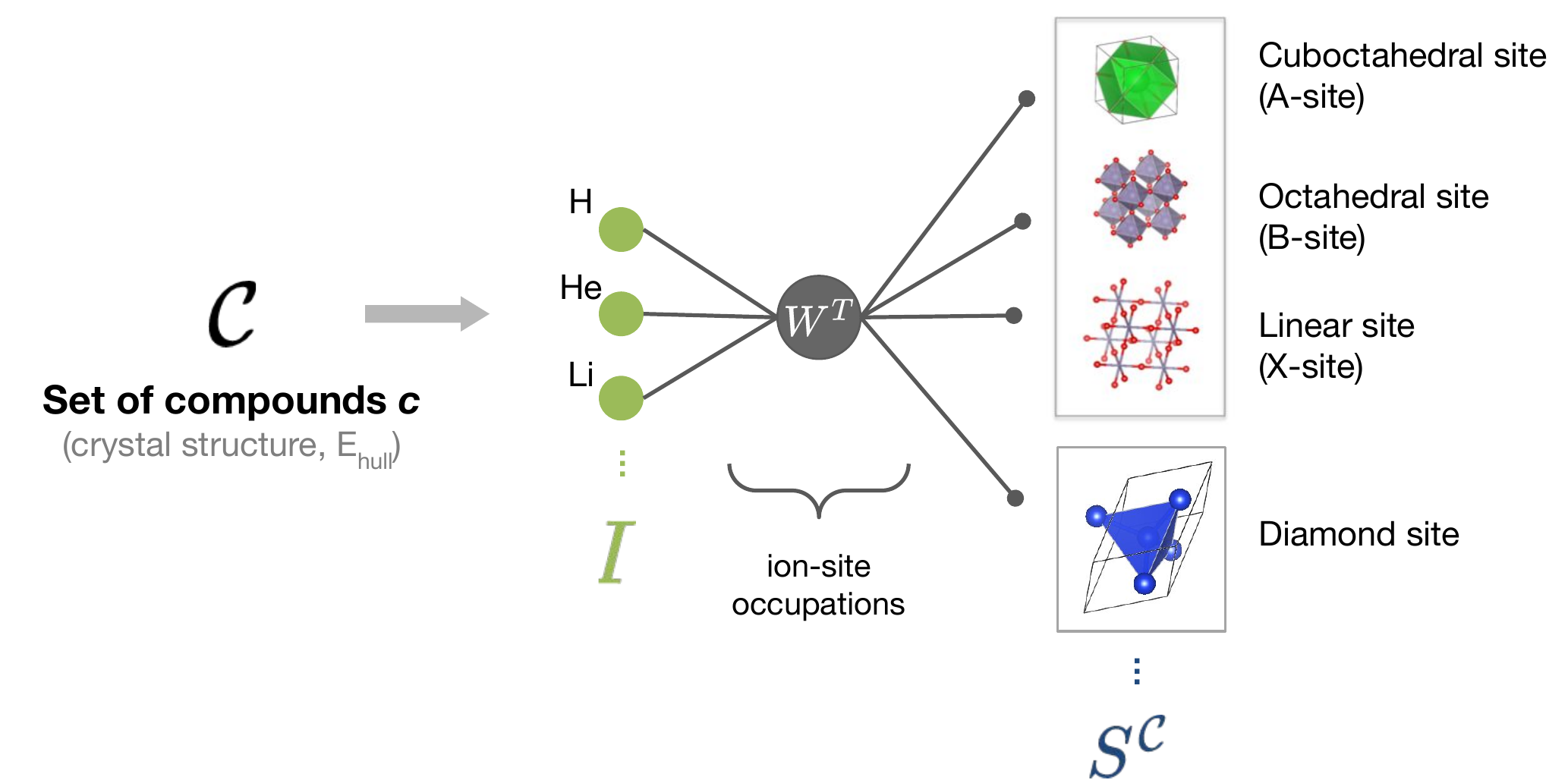}
    \caption{Illustration of the bipartite graph $G^T = (I, S^\mathcal{C}, E^+, W^T$)}.
    \label{fig:graph}
\end{figure*}

Figure~\ref{fig:graph} pictorially illustrates the structure of the bipartite graph $G^T$ built from the set of compounds $\mathcal{C}$.

\subsubsection{The Embedding for Ions and AM sites}

In order to generate embeddings that are encoded into vector representations for each type of node (ions and AM sites) we conduct random walks across the graph. DeepWalk~\cite{deepwalk2014}, a widely-used algorithm, generates these random walks by starting at a randomly selected node and choosing the next node uniformly at random from its neighbors. This process is repeated for a fixed number of steps to generate a single random walk, and is performed multiple times to generate a set of random walks for the entire graph. The premise of DeepWalk is that nodes frequently co-occurring on these walks likely have similar roles or positions in the graph, hence the learned embeddings will capture these similarities. Once the random walks are generated, they are input into the Word2Vec~\cite{word2vec2013} algorithm to learn the embeddings. These embeddings, which capture the local and global structure of the graph, can then be used for downstream tasks such as link prediction or node classification.

This methodology was adapted for a weighted graph. A random walk $rw$ in the weighted bipartite graph $G$ can be defined as a sequence of alternating vertices from the sets $I$ and $S^\mathcal{C}$, where the probability of transitioning from one vertex to another depends on the normalized weights of the edges $W^T(i, s)$ linking to the current graph node.

First, we calculate the transition probabilities for each ion and AM site at parametric temperature $T$. For each ion $i \in I$ and each AM site $s \in S^{\mathcal{C}}$, the sum of weights $W^T(i, s)$ for all edges connected to it is computed. Then, the transition probability $P(i, s)$ is given as follows:



\begin{equation}
P(i, s) = \frac{W^T(i, s)}{\sum_{s \in S^{\mathcal{C}}} W^T(i, s)}
\end{equation}

A single $rw$'s in $G$ can be expressed as a sequence of alternating vertices from $I$ and $S$, $rw = (i_1, s_1, i_2, s_2, \dots, i_n, s_n)$, where $i_k \in I$, $s_k \in S^{\mathcal{C}}$, and the choice of $s_k$ given $i_k$ is determined by the transition probability $P(i_k, s_k)$, and the choice of $i_{k+1}$ given $s_k$ is determined by the transition probability $P(s_k, i_{k+1})$. A total of 150 random walks per node, with a walk length of 50 nodes was used.

The set of all $rw$ in the graph is defined as $RW$. The Word2Vec function takes $RW$ as input and a set of hyperparameters $\theta$ and outputs an embedding function $\phi_\theta=\text{Word2Vec}(RW, \theta)$.

\begin{figure*}[ht]
    \centering
    \includegraphics[width=\textwidth]{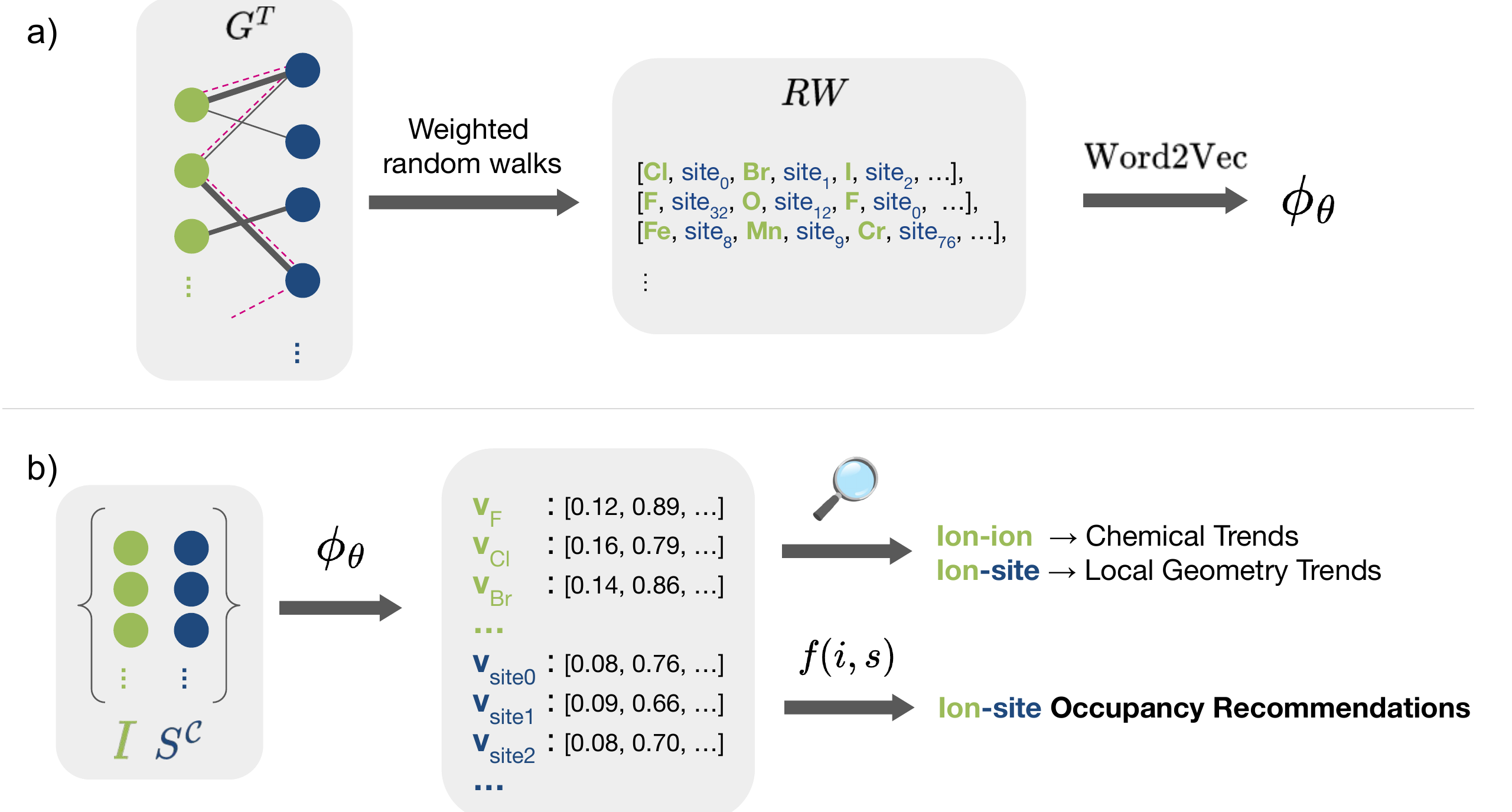}
    \caption{a) Graphical representation of the embedding creation workflow:  given a graph $G^T$, the set of weighted random walks is made and then used as input for Word2Vec. The result is an embedding function $\phi_theta$. b) Results obtained from the embedding's vector representations of ions and AM sites.}
    \label{fig:graph_workflow}
\end{figure*}

The Figure \ref{fig:graph_workflow}a illustrates the embedding creation process.

\subsubsection{Defining the function $f$}

Our RS is encoded in the function $f(i,s)$. The embedding function $\phi_\theta$ maps a node (either ion $i \in I$ or a AM site $s \in S$) to a $d$-dimensional vector representation. These representations are then processed by a similarity function, with the output fed into a binary classification model. We employ cosine similarity as the similarity metric and a decision tree (DT) as the model. 

For training, we adapted the dataset $\mathcal{D}$ creation methodology for weighted graphs. Edges $E^+$ were sampled by their weight, with higher weight edges having a greater selection chance. Once selected, the edge was removed from the original $G^T$ graph to prevent data leakage during embedding. The edge's weight was then used as a multiplier for instance repetition in the dataset. As an example, a positive edge with a weight of 10 would be repeated 10 times. Negative examples were uniformly sampled from $E^-$ until an approximate 50\% ratio of positive and negative examples was achieved.

The dataset was divided into 80\% for training, 10\% for validation, and 10\% for testing. Optimal hyperparameters for each temperature were determined using the Optuna python library~\cite{optuna2019}. We optimized the hyperparameters $\theta$ of the embedding function $\phi$, as these have been shown to significantly impact recommender systems~\cite{word2vec_hp2018}. The accuracy results for the optimal hyperparameters are presented in the Supplementary Information.

After training, the learned function $f$ was used to predict the likelihood of edges $(i, s) \in E^-$, between node pairs in the graph, and then use these new reccomended occupations for building new compounds.

\section{Results and Discussion}

Establishing an embedding that encapsulates the structure of bipartite graphs $G^T$ facilitated two distinct analyses concerning the relations among the graph entities. The first one pertains to the relationships between ions, while the second one pertains to the relationship between the ions and the AM sites.

The first was done by correlating the ion-ion embedding cosine similarities to ionic substitution in crystal structures. This approach enabled the validation of chemical trends within the continuous embedding space created solely with ion-site occupation data.

The second analysis involved the study of proximate AM sites surrounding the ions within the embedding space. This approach produced two distinct types of outcomes. Firstly, it enabled the verification of the overarching local geometries of the AM sites in association with the corresponding ion. Secondly, it allowed the exploration of novel ion-site occupancies informed by the trained RS function $f$. 

Figure \ref{fig:graph_workflow}b illustrates the results obtained from the embedding and further discussed below.

\subsection{Embedding Chemical Trends}

\begin{figure}
    \centering
    \includegraphics[width=\textwidth]{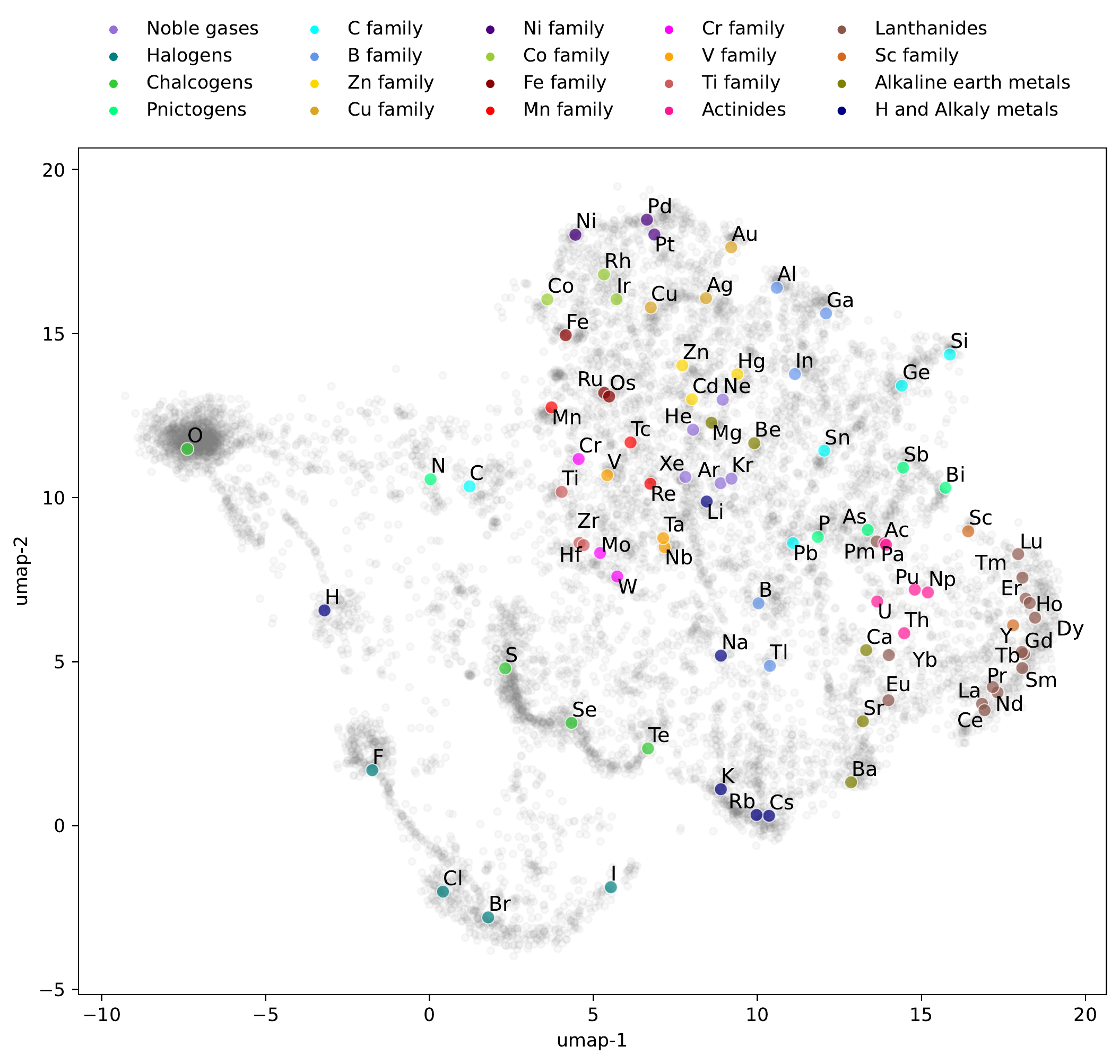}
    \caption{Embedding of $G^0$ built with $\text{OQMD}^{\text{current}}$. In light gray are the AM sites. The colored points represent the set of 89 ions, each color representing its corresponding family in the periodic table. UMAP parameters: metric: cosine, n\_neighbours: 250, min\_dist: 0.5, densmap: True, random\_state: 0.}
    \label{fig:embedding}
\end{figure}

After optimizing the hyperparameters of Word2Vec, an embedding was obtained from $G^0$,  exclusively considering compounds on the hull and built with $\text{OQMD}^{\text{current}}$. We applied the uniform manifold approximation and projection (UMAP) technique to reduce the dimensionality of the embedding~\cite{umap} for a two-dimensional visualization (Figure~\ref{fig:embedding}).

The embedding shows that, with some few distinct exceptions, the ions from the same family in the periodic table usually occupy the same group of sites, as they are close together in the embedding space projection of Figure~\ref{fig:embedding}. Within the same cluster family, there is also a larger proximity between ions with similar atomic radii. For instance, within the halogen family cluster, Chlorine (Cl) is more closely related to Bromine (Br) than to Iodine (I).

A similar pattern is also observed within transition metal clusters. Ions from the 5d period display a greater likeness to 4d period ions than those from the 3d period. This can be exemplified by 5d transition metals (Hf, Ta, W, Re, Os, Ir, Pt), which showcase similar atomic radii to 4d metals within the same group (Zr, Nb, Mo, Tc, Ru, Rh, Pd). This phenomena can be explained by the lanthanide contraction effect\cite{Housecroft_Sharpe_2018}. As one moves through the lanthanide series (f-block elements) on the periodic table, the atomic radii tend to decrease due to the poor shielding of the increasing nuclear charge by the 4f electrons, resulting in a higher effective nuclear charge and a smaller atomic radius. This effect continues into the subsequent d-block elements, causing the 5d elements to have atomic radii similar to their 4d counterparts, despite being in a new period. 

The embedding projection also illuminates trends related to the ionic character of ions. Anions, such as oxygen, nitrogen, carbon, hydrogen, halogens, and chalcogens, largely congregate on the left side, while cations typically occupy the right side. This pattern supports the notion that cations and anions tend to inhabit distinct sites in crystal structures.

The visualization also underscores a sizable and dense cluster surrounding oxygen (O). This clustering likely stems from the original ICSD~\cite{ICSD} database that serves as the foundation for OQMD. ICSD primarily focuses on published crystallographic research, which heavily features oxides due to their prevalence and to their wide-ranging applications.

A unique trend is apparent with ions in the second period, or the first row of p-block elements (B, C, N, O, F), as these elements behave differently from their same-family heavier counterparts. This divergence arises from their increased electronegativity and smaller atomic radii, affecting bond polarity and lengths.

Certain ion-ion similarities spanning across different periodic families can be observed, often due to shared oxidation states. For instance, Ytterbium (Yb) and Europium (Eu) from the lanthanide series show a striking resemblance to alkaline earth metals, likely attributable to their shared +2 oxidation state. Thallium (Tl), despite belonging to a different group, demonstrates notable proximity to Sodium (Na) in the embedding, an occurrence that can be explained by Thallium's ability to adopt a +1 oxidation state. In addition, Scandium (Sc) and Yttrium (Y), which typically present a +3 oxidation state, are found to be closely aligned with the lanthanides, further emphasizing the influence of overlapping oxidation states on ion-ion relationships.

Supplementary to the UMAP visualization, a dendrogram depicting the hierarchical clustering of the 89 ions within the embedding space is provided in the Supplementary Information.

\subsection{Embedding Local Geometry Trends}

In further analysis, we generated heatmaps to provide insights into the local geometric configurations of the sites. Utilizing the CrystalNN method\cite{zimmermann2020local, crystalnn2021}, we computed scores representing the likelihood of each type of local geometry for every site. Subsequently, we divided the 2D UMAP space and computed the average score for each segment. To smooth out the data distribution, we applied a Gaussian fitting across the 2D space, focusing on the dimension representing the geometry of interest. 

\begin{figure}
\centering
\includegraphics[width=\textwidth]{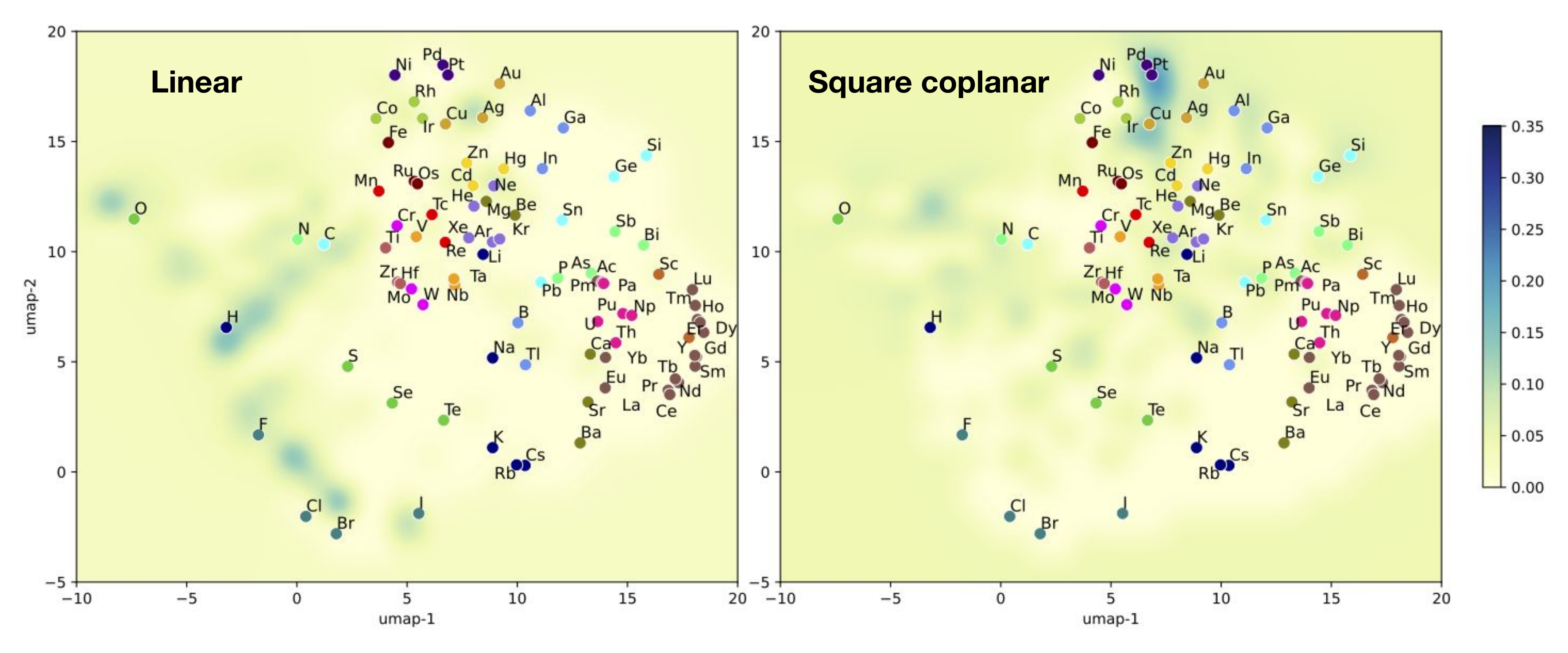}
\caption{Heatmaps indicating local geometry coordination for linear and square coplanar configurations. The color scale corresponds to the density of sites within the given area of the UMAP 2D-transformed embedding space that conform to the respective geometry.}
\label{fig:heatmaps}
\end{figure}

The results for two distinct types of local geometries—linear and square coplanar—are presented in Figure~\ref{fig:heatmaps}. The color variations surrounding the ions in these maps correspond to the degree to which ions occupy stable compound's sites with the given geometry.

Something important to note is that, as the color encodes a site average over the area, a light color does not necessarily imply that the ion is not capable of occupying a site with the given local geometry. It implies instead that, statistically in the data set, that geometry is less common than other geometries.

The heatmap for linear local geometry shows that sites with this geometry tend to concentrate around anions such as halogens, H and O. Cu and Ag also present a certain concentration around them, which could be attributed to their capacity of having a less common oxidation state of 1+. An example of such occupation is found in the delaffosite class of compounds~\cite{delafossite1971}.

The square planar coordination is not as common as other geometries but is typically found with certain transition metals that present a 2+ oxidation state. For instance, in certain crystal structures, Pd, Pt, and Cu can adopt square planar geometry. The heatmap confirms this and suggests that Ir and Zn, which can also present 2+ oxidation state, can also adopt this geometry. One prominent such family is that of the copper oxide high-temperature superconductors \cite{Bednorz_ZPB:1986,Proust_ARCMP:2019}, where the square planar geometry and the Cu 2+ oxidation state are believed to be two crucial ingredients for the distinctive properties of these materials

The heatmaps for other local geometries—cuboctahedral, octahedral, tetrahedral, trigonal planar, and single bound—are presented in the Supplementary Information.

\subsection{Ion-site Recommendations}

\subsubsection{Historical Validation on Cubic Halide Perovskites}

Utilizing the methodology outlined in the previous section, we trained a RS using $\mathcal{C} = \text{OQMD}^{\text{past}}$. We then validated the methodology using the set of new cubic halide perovskites in $(\text{OQMD}^{\text{current}}-\text{OQMD}^{\text{past}})$. Our objective was to assess the impact of the parametric temperature $T$ on two key aspects of RSs: the degree to which it restricts the search space; and the number of thermodynamically stable compounds from $(\text{OQMD}^{\text{current}}-\text{OQMD}^{\text{past}})$ that can be identified within this restricted search space by our framework.

Halide perovskites are important since they are a promising class of materials for optoelectronics and photovoltaics.\cite{Walsh2021} They possess unique crystal structures and exceptional optoelectronic properties, including high absorption coefficients and long charge carrier diffusion lengths. Halide perovskite solar cells have achieved impressive power conversion efficiencies, surpassing traditional thin-film technologies.\cite{Tian2020} These materials can be fabricated using low-cost techniques and integrated into various devices such as LEDs and photodetectors. However, challenges remain in their stability, and ongoing research aims to improve their performance and understand their fundamental properties.

A comparison to a brute-force approach needs to be done when evaluating a RS. A cubic halide perovskites brute-force search has a space size of 11,200 possibilities, provided we consider four halogens \{F, Cl, Br, I\}, only non-metalic ions, and ions from the 1st and 2nd row for the A and B sites, while excluding noble gases, radioactive ions Additionally, for the A site, we eliminated the option of p-orbital ions. 

$\text{OQMD}^{\text{past}}$ has a total of 135 cubic halide perovskites, 63 of which are thermodynamically stable ($E_{\text{hull}} \leq 30 \text{meV/atom}$). $\text{OQMD}^{\text{current}}$ has 432 additional cubic halide perovskites, 50 of which are thermodynamically stable. In the context of a binary classification problem, these 50 compounds constitute the set of positive instances, while the remaining 382 instances are defined as the negative class of new thermodynamically unstable compounds. Since the positive and negative sets are unbalanced and false positives are undesired in the context of a RS, the precision was the chosen metric to evaluate this task. Precision is a measure of relevancy — it quantifies the proportion of recommended items that are actually relevant to the user. The results for parametric temperatures of $T = \{0K, 100K, 300K\}$ are presented in Table~\ref{table:temp_results}.

\begin{table}[ht]
\centering
\caption{Comparison of different models with different temperatures to the brute-force approach. The search space is the number of possible combinations given the ion-site suggestions made by the RS. The precision is given within $(\text{OQMD}^{\text{current}} - \text{OQMD}^{\text{past}})$ by the ratio of the thermodynamically stable compounds (true positives) to the total number of compounds suggested by the recommender (true positives + false positives).}

\begin{tabular}{l c c c c}
\toprule
 & $G^{0}$ & $G^{100}$ & $G^{300}$ & Brute-force \\ \midrule
 A search space& 9 & 9 & 12 & 50 \\
 B search space & 11 & 12 & 15 & 57 \\
\ce{ABX3} search space & 392 (3.5\%) & 428 (3.8\%) & 708 (6.3\%) & 11,200 (100\%) \\
In $(\text{OQMD}^{\text{current}} - \text{OQMD}^{\text{past}})$ & 49 & 73 & 109 & 432 \\
$E_{\text{hull}} \leq 30\text{ meV/atom}$ & 13 & 23 & 28 & 50 \\
Precision & 26.5\% & \textbf{31.5\%} & 25.7\% & 11.6\% \\ 
\bottomrule
\end{tabular}
\label{table:temp_results}
\end{table}

The results indicate that $G^{100}$ provides the optimal balance between restricting the search space and offering the largest proportion of thermodynamically stable recommendations. These findings suggest that including all compounds and weighting them by their $E_{\text{hull}}$ is the most effective strategy. The comparatively poorer performance of $T = 300K$ suggests that a higher temperature underpenalizes non-stable compounds.

Another trend that can be attributed to the parametric temperature is on the \ce{ABX3} search space reduction: a higher $T$ resulted in a higher number of possibilites. Such result can be attributed to the higher connectivity of ions to AM sites in the resulting $G^T$ graphs, resulting from $W^T(i,s)$.

It is worth noting that there are compounds that were recommended that are not present in $(\text{OQMD}^{\text{current}} - \text{OQMD}^{\text{past}})$ and thus remain to be verified with DFT calculations. The list of the recommendations for A and B sites are given in the Supplementary Information.

\subsubsection{Recommendations for Kagome Lattice Prototype Structure}
\label{kagomeprediction}

Using the validated methodology, optimal hyperparameters~$\theta$, and the best parametric temperature $T=100K$, we trained a recommender system with $\text{OQMD}^{\text{current}}$. This was then used to recommend ion-site occupations for the sites of the class of compounds \ce{AB3C5}.

Their crystal structure presents a layered structure of Kagome lattices separated by A-ions in a honeycomb lattice. The Kagome lattice is made of three B-ions coordinated with one C-ion, sandwiched between two honeycomb lattices of C. The structure presents a symmetry given by the space group $P6/mmm$. 

Ortiz et al first synthesized this prototype structure in 2019 with the compositions \ce{KV3Sb5} , \ce{RbV3Sb5}, \ce{CsV3Sb5}\cite{kagome2019}. Since then, many characterizations of these compounds were made, revealing properties such as chiral charge ordering~\cite{jiang2021unconventional}, large anomalous Hall effect~\cite{hall_kagome2020}, non-trivial topological band structures and superconducting ground states~\cite{ortiz2021superconductivity, csv3sb5_2020}.

\begin{figure}
    \centering
    \includegraphics[width=13cm]{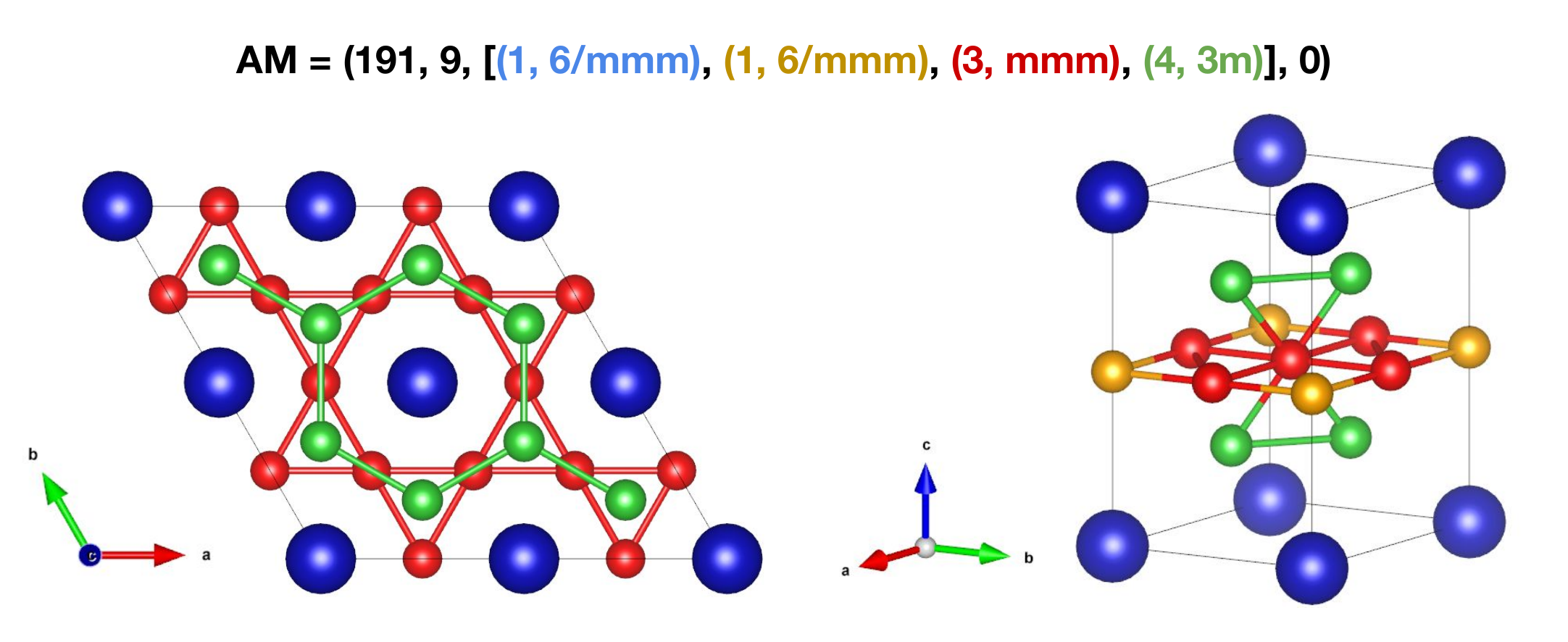}
    \caption{Prototype Crystal structure of compound \ce{CsV3Sb5}, with stoichiometry \ce{AB3C5}. The ions A and B are represented in blue and red, respectively. The C ion occupies two non-equivalent types of sites, represented in green and in yellow.}
    \label{fig:kagome}
\end{figure}

\begin{sloppypar}
The prototype crystal structure maps to the four non-equivalent sites $(191, 9, [(1, 6/mmm), (1, 6/mmm), (3, mmm), (4, 3m)], 0)$ AM label, as shown in Figure~\ref{fig:kagome}. In this prototype, A occupies the first site in the $Wy$ list $(1, 6/mmm)$, B occupies the third site $(3, mmm)$, and C occupies both the second and the fourth sites $(1, 6/mmm)$ and $(4, 3m)$. For simplicity, we truncated the recommendations to a \ce{AB3C5} stoichiometry, using the intersection of recommendations of C for both sites (represented by yellow and green atoms in Figure~\ref{fig:kagome}). 
\end{sloppypar}

We input these sites into the RS, which recommended ions to occupy them. Known compositions for this structure in $\text{OQMD}^{\text{current}}$ were filtered out, and the remaining recommendations were ranked according to the sum of ion-site distances in the embedding space. The top 10 recommendations were then verified using DFT calculations, and the results are presented in Table \ref{table:rs_results}. All ion-site recommendations are given in the Supplementary Information.

\begin{table}[ht]
\centering
\caption{Thermodynamical stability results for the top 10 compounds built with the ion-site suggestions of the RS built with T = 100K and $\text{OQMD}^{\text{current}}$.}
\begin{tabular}{c c c r c c}
\toprule
A & B & C & Composition & $\Sigma$(ion-site distances) & \shortstack{$E_{\text{hull}}$ \\(meV/atom)} \\ \midrule
Rb & Ti & Sb & \textbf{\ce{RbTi3Sb5}} & 0.125 & 0 \\
K & Ti & Sb & \textbf{\ce{KTi3Sb5}} & 0.128 & 13 \\
Cs & Ti & Sb & \textbf{\ce{CsTi3Sb5}} & 0.134 & 0 \\
Rb & Ti & Bi & \textbf{\ce{RbTi3Bi5}} & 0.173 & 5 \\
K & Ti & Bi & \textbf{\ce{KTi3Bi5}} & 0.176 & 23 \\
Cs & Ti & Bi & \textbf{\ce{CsTi3Bi5}} & 0.182 & 0 \\
Rb & Hf & Sb & \ce{RbHf3Sb5} & 0.219 & 182 \\
K & Hf & Sb & \ce{KHf3Sb5} & 0.223 & 165 \\
Tl & Ti & Sb & \ce{TlTi3Sb5} & 0.225 & 102 \\
Cs & Hf & Sb & \ce{CsHf3Sb5} & 0.228 & 139 \\ \bottomrule
\end{tabular}
\label{table:rs_results}
\end{table}

The results show that 6 out of the 10 recommendations are indeed stable (with $E_{\text{hull}} < 30 \text{ meV/atom}$). These 6 compounds were also reported by Jiang et al~\cite{screening_kagome2022} as having a $E_{\text{hull}} < 30 \text{ meV/atom}$, but at the cost of a total of 1386 high-throughput DFT calculations performed. Notably, these materials were entirely new in $\text{OQMD}^{\text{current}}$. 

Another significant finding is the novel recommendation of the ion family Ti, Hf, and Zr for the B site occupancy. This type of occupation was not previously present in $\text{OQMD}^{\text{current}}$, making it a non-trivial recommendation that could not have been made simply by visually or intuitively filling the gaps in the periodic table.

\section{Conclusion}

Leveraging data from OQMD, we successfully constructed a graph-based recommender system that incorporates information on the thermodynamic stability of compounds along with symmetry and local geometry of sites. The developed embedding enabled us to discern chemical and local geometrical trends through ion-ion similarities, and to extract ion-site recommendations from the ion-site relationships.

Our experiments with a parametric temperature that weights the $E_{\text{hull}}$ of compounds led us to conclude that considering not only the compounds on the hull, but also those above the hull, can be beneficial \ to the identification of new compounds.

By employing a recommender system with optimally tuned hyperparameters and parametric temperature, we managed to propose novel compounds that are not present in the current version of OQMD. This demonstrates the potential of our approach to facilitate the discovery of new materials.

\section{Data Availability}

The RS trained with $\text{OQMD}^{\text{current}}$, along with the graphs $G^{100}$ and $G^0$ (as the data used to build them), as well as the ion and site embeddings, can be downloaded from our GitHub repository at \url{https://github.com/simcomat/ionic-sub-RS}. A detailed explanation of the process for obtaining recommendations for ion-site occupations and new compounds is provided in a Jupyter notebook available on the same page.

\section{Competing Interests}

The Authors declare no Competing Financial or Non-Financial Interests.

\section{Author Contributions}

E.O. conceived the work, built the model, and ran the DFT calculations and wrote the first version of the paper. H.F., J.N.B.R. and G.M.D. helped with refinements of the model. G.M.D. supervised the work. All authors contributed with the final version of the paper.

\begin{acknowledgement}

The authors acknowledge the National Laboratory for Scientific Computing (LNCC/MCTI, Brazil) for providing HPC resources of the SDumont supercomputer. The authors also thank the Brazilian funding agencies FAPESP (grants 17/02317-2 and 18/11641-0), CNPq (309384/2020-6) and CAPES for financial support. This research has been conducted within the scope of the Materials Informatics INCT (CNPq). 

\end{acknowledgement}

\bibliography{achemso-demo}

\end{document}


\section{DFT Calculations}

All Density Functional Theory (DFT) calculations were performed using the Vienna Ab-initio Simulation Package (VASP) v6.3.2. Electron exchange and correlation were depicted using Perdew, Burke, and Ernzerhof's (PBE) generalized gradient approximation (GGA), with potentials generated via the projected augmented wave (PAW) method. Convergence thresholds were set at 0.0001 eV/atom for electronic ground states and at 0.001 eV/atom for ionic steps. Consistent with the OQMD's calculations, we adopted a plane wave kinetic energy cutoff of 520 eV.

The calculation for energy above hull $E_\text{hull}$, which represents the difference in energy between a particular compound and the lowest energy compound or set of phase-separated compounds at a corresponding composition, was accomplished by obtaining the total energies of all existing compounds within the composition system from the Open Quantum Materials Database (OQMD).

\section{Hyperparameter Optimization}

Utilizing the optimization framework Optuna\cite{optuna2019}, we explored the following hyperparameters of the set $\theta$  of Word2Vec, implemented by gensim\cite{gensim2010}: vector size, window size, negative, ns exponent, and sampling. The search spaces for each hyperparameter are given in Table \ref{table:search_space}. We utilized the skip-gram model version of Word2Vec.

\begin{table}[h]
\centering
\caption{Search spaces of Word2Vec's explored hyperparameters.}
\begin{tabular}{l l}
\toprule
Word2Vec hyperparameters & Search space \\ \midrule
window & [2, 3, 4, 5, 6] \\ 
negative & [5, 10, 15] \\
ns exponent & [-1.0, -0.75, -0.5, -0.25, 0.0, 0.25, 0.5, 0.75, 1.0] \\ 
sample & [1e-5, 1e-4, 1e-3, 1e-2, 1e-1] \\ 
vector size & [4, 8, 16, 32, 64] \\ \bottomrule
\end{tabular}
\label{table:search_space}
\end{table}

The optimization in the linkage prediction task was performed for temperatures $T = \{0K, 100K, 300K\}$ with $\mathcal{C} = \text{OQMD}^{\text{past}}$, and with temperature $T = 100K$ for $\mathcal{C} = \text{OQMD}^{\text{current}}$. The results are presented in the Table \ref{table:optimization}.

\begin{table}[h]
\centering
\caption{Optimal hyperparameters for the link prediction task.}
\begin{tabular}{l c c c c c c c}
 \toprule
$\text{OQMD}^{\text{past}}$ & negative & ns exponent & sample & vector size & window & Accuracy \\ \midrule
$G^0$ & 15 &  0.25 & 0.00001 & 16 & 3 & 83.7\% \\
$G^{100}$ & 5 & -0.75 & 0.01 & 16 & 4 & 84.4\% \\
$G^{300}$ & 5 & 0.00 & 0.01 & 16 & 5 & 84.1\% \\ \midrule
$\text{OQMD}^{\text{current}}$ \\ \hline
$G^{0}$ & 5 &  0.75 & 0.001 & 8 & 5 & 84.8\% \\ 
$G^{100}$ & 10 & -0.25 & 0.1 & 8 & 5 & 85.1\% \\
\bottomrule
\end{tabular}
\label{table:optimization}
\end{table}

\section{Ion Dendogram}

\begin{figure}[h]
    \centering
    \includegraphics[height=\textheight-1.5cm]{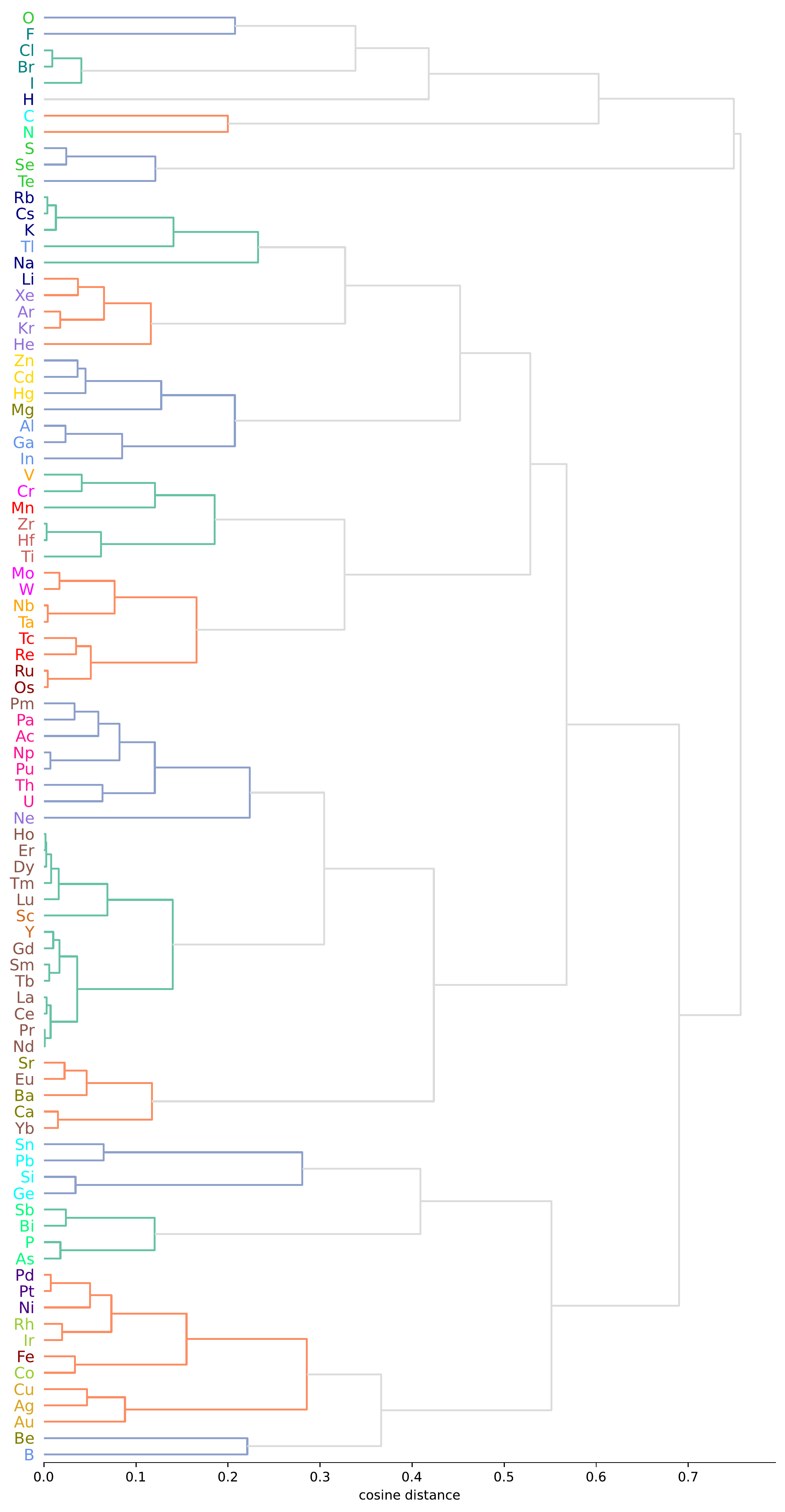}
    \caption{Dendogram built with the hierarchical clustering methodology. Linkage method: weighted, metric: cosine, color\_threshold: 0.3.}
    \label{fig:dendogram}
\end{figure}

An alternative way of visualizing the embedding space, more specifically the ion-ion relationships, is to use the hierarchical clustering methodology. The Figure~\ref{fig:dendogram} shows the resulting dendogram. 

\section{Local Geometry Heatmaps}

In this section we present the heatmaps of local geometry of sites for the following: single bound, trigonal planar, tetrahedral, octahedral and cuboctahedral.

\begin{figure}[h]
    \centering
    \includegraphics[width=\textwidth]{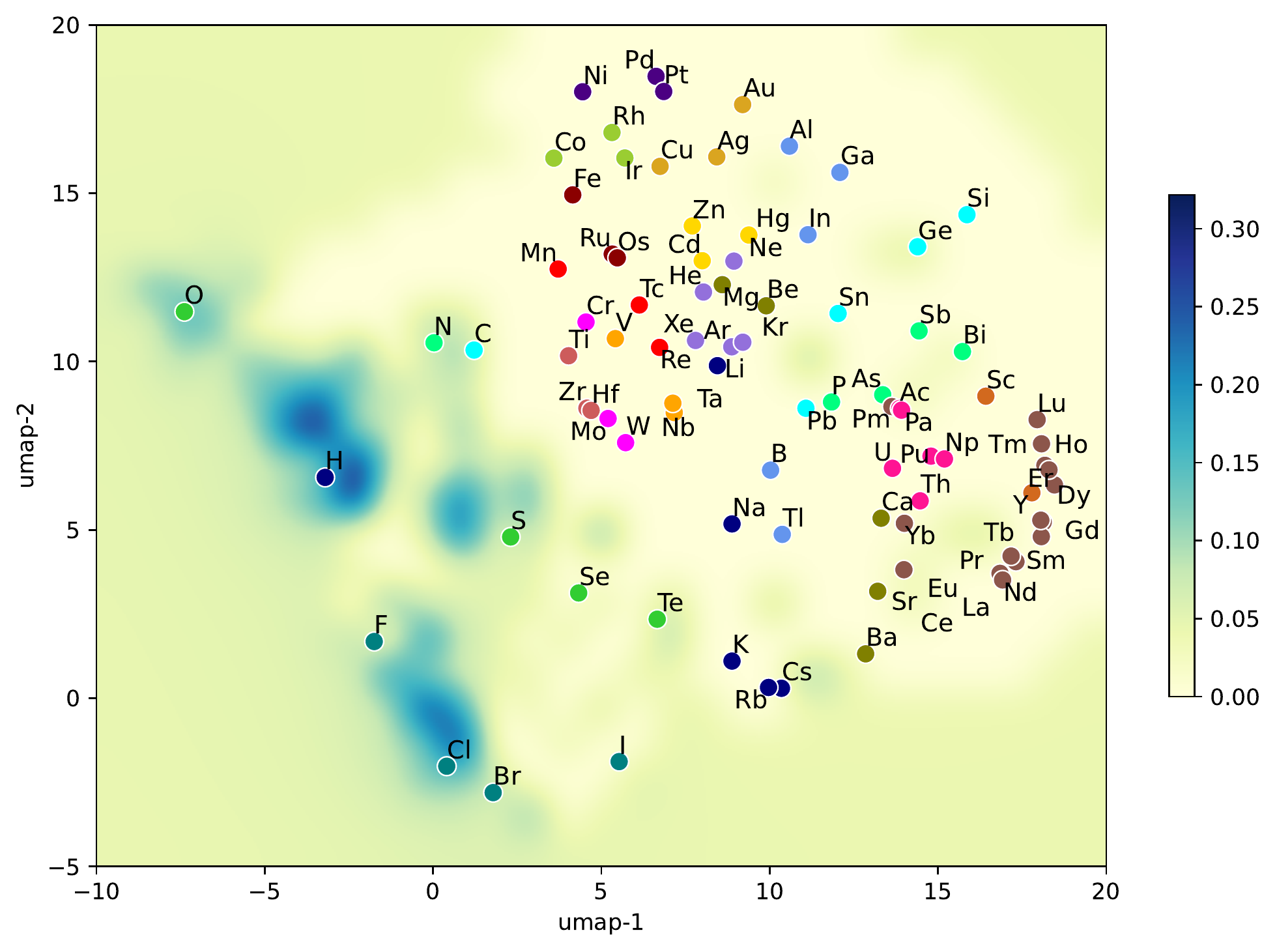}
    \caption{Single bound coordination environment heatmap}
\end{figure}

\begin{figure}[h]
    \centering
    \includegraphics[width=\textwidth]{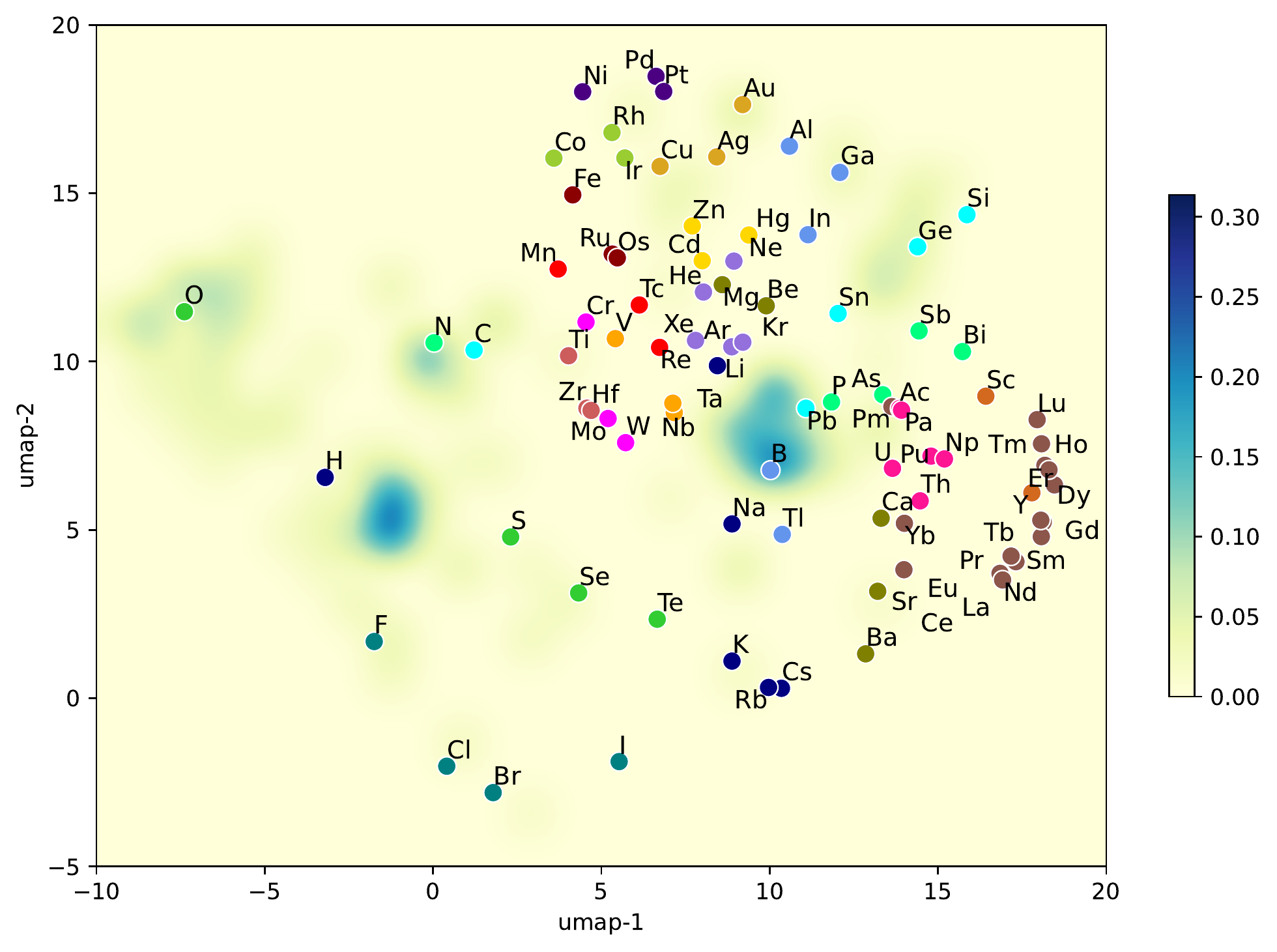}
    \caption{Trigonal planar coordination environment heatmap}
\end{figure}

\begin{figure}[h]
    \centering
    \includegraphics[width=\textwidth]{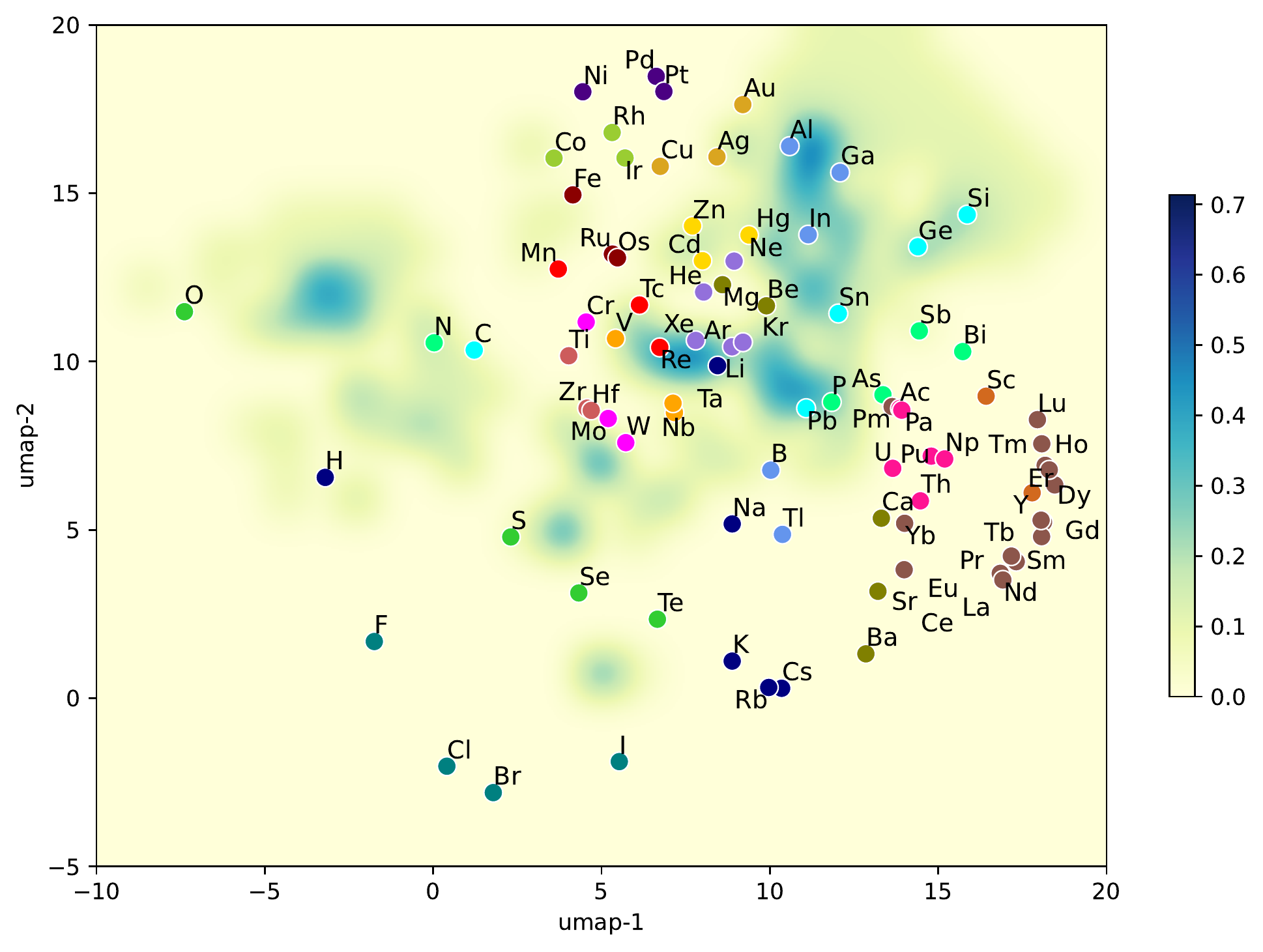}
    \caption{Tetrahedral coordination environment heatmap}
\end{figure}

\begin{figure}[h]
    \centering
    \includegraphics[width=\textwidth]{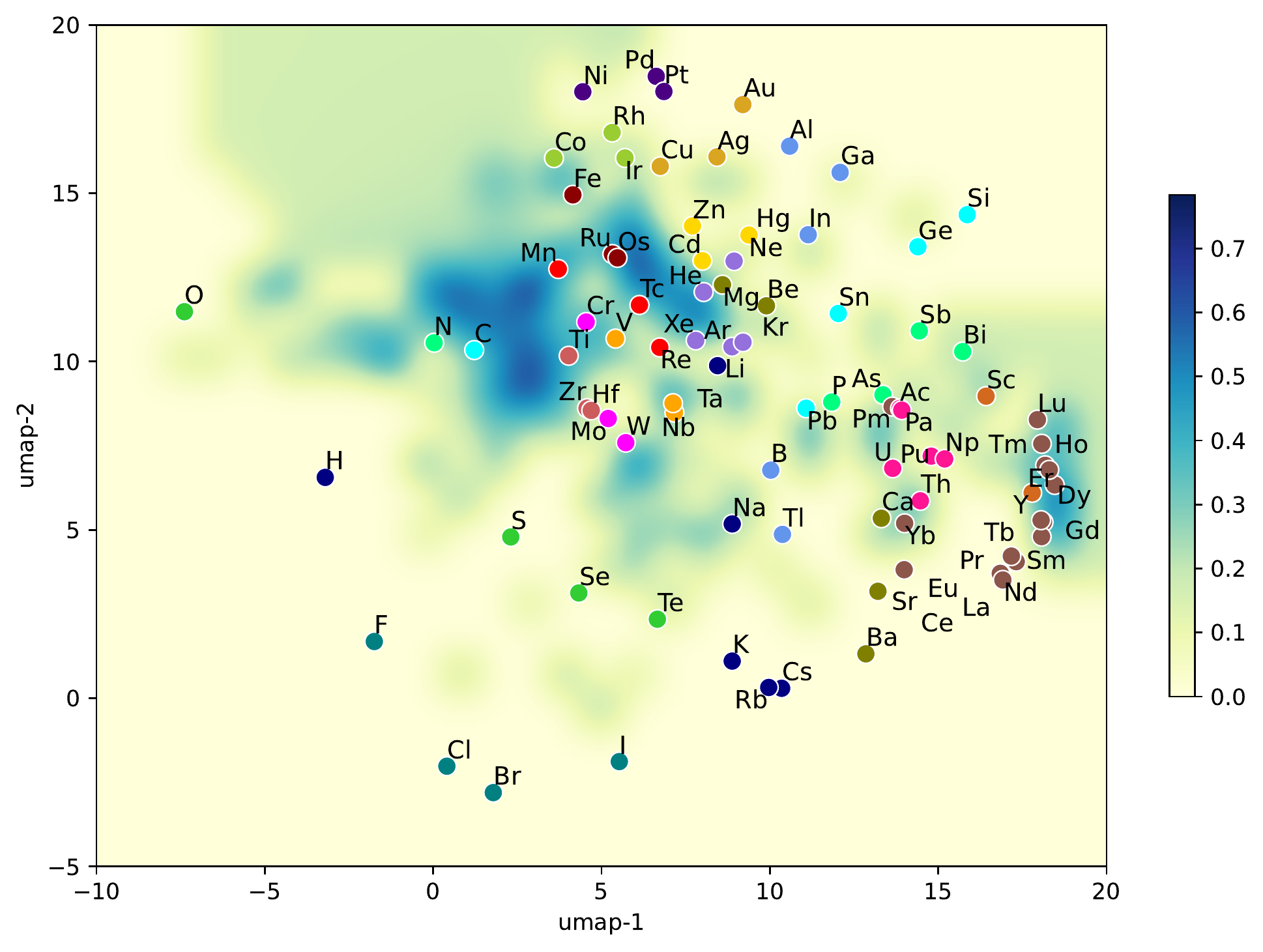}
    \caption{Octahedral coordination environment heatmap}
\end{figure}

\begin{figure}[h]
    \centering
    \includegraphics[width=\textwidth]{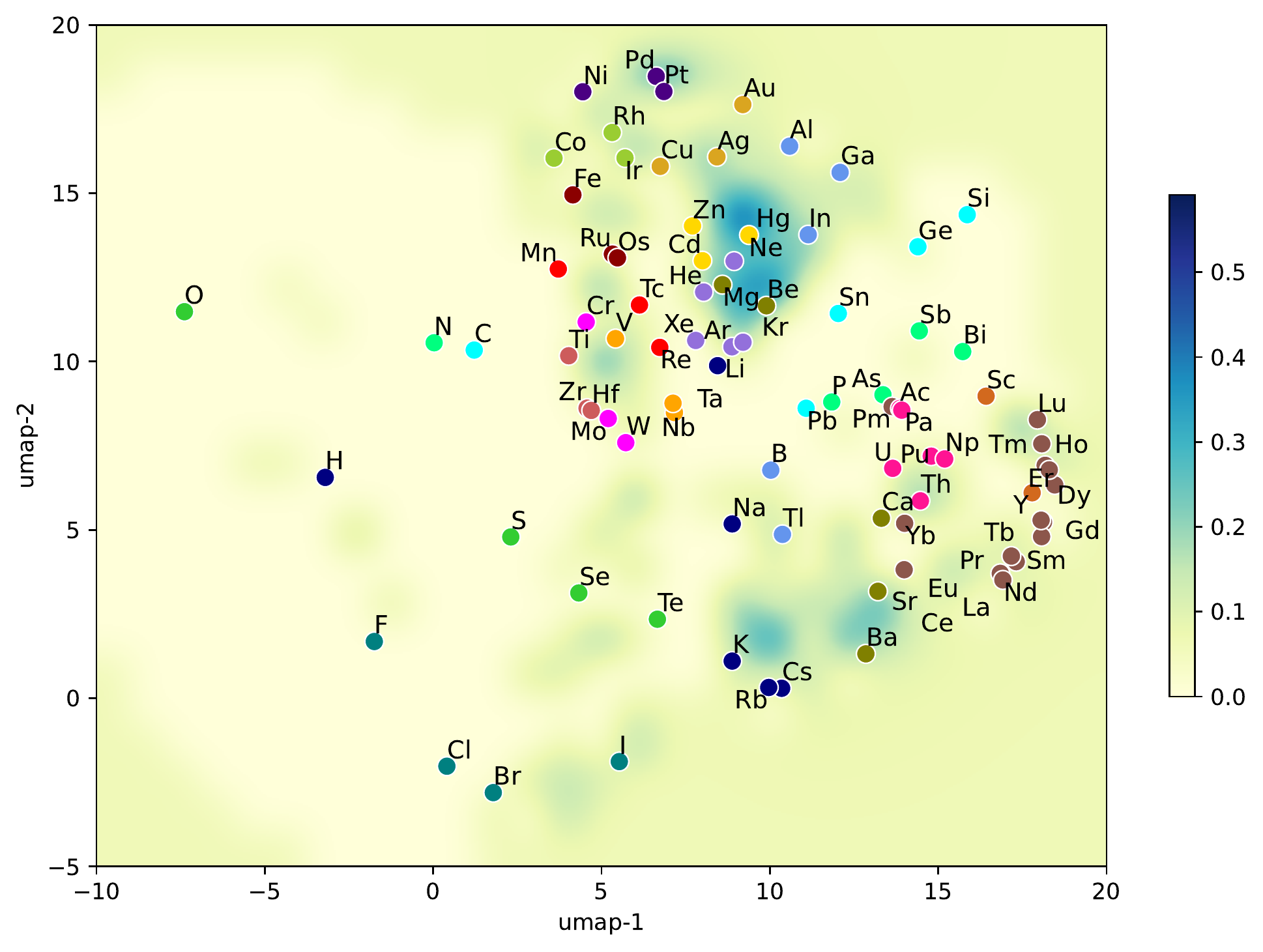}
    \caption{Cuboctahedral coordination environment heatmap}
\end{figure}

\section{Cubic Halide Perovskites Recommendations}

The Table~\ref{table:perovskite_rec} shows all ion-site recommendations given for the A and B sites by the RS system built from the optimized $G^{T}$, and  with $\mathcal{C} = \text{OQMD}^{\text{past}}$.

\begin{table}[ht]
\centering
\caption{Recommendations for the sites A and B of cubic halide perovskites, ordered by ion-site distances in the embedding space.}
\begin{tabular}{c c c c c}
\toprule
 & \shortstack{ion-site dist \\threshold} & \shortstack{A site \\(ion-site dist)} & \shortstack{B site \\(ion-site dist)} & \shortstack{\# of \ce{ABX3} \\combinations} \\ \midrule
$G^0$ & 0.417 & \begin{tabular}[]{@{}c@{}}Rb (0.067)\\ Cs (0.073)\\ K (0.093)\\ Ba (0.185)\\ Sr (0.273)\\ Eu (0.304)\\ Na (0.355)\\ Ca (0.366)\\ Yb (0.398)\end{tabular} & \begin{tabular}[]{@{}c@{}}Os (0.276)\\ Ru (0.292)\\ Sc (0.346)\\ Mg (0.350)\\ Ag (0.353)\\ Rh (0.358)\\ Ir (0.361)\\ W (0.388)\\ Re (0.392)\\ Cd (0.403)\\ Na (0.409)\end{tabular} & 392 \\ \midrule
$G^{100}$ & 0.410 & \begin{tabular}[]{@{}c@{}}Cs (0.091)\\ Rb (0.124)\\ K (0.164)\\ Ba (0.174)\\ Sr (0.271)\\ Eu (0.295)\\ Na (0.348)\\ Ca (0.353)\\ Yb (0.39)\end{tabular} & \begin{tabular}[]{@{}c@{}}Os (0.279)\\ Ru (0.296)\\ Ag (0.296)\\ Na (0.329)\\ Re (0.340)\\ Cu (0.349)\\ Rh (0.372)\\ Ir (0.374)\\ Sc (0.382)\\ Mg (0.399)\\ Cr (0.404)\\ Mn (0.410)\end{tabular} & 428 \\ \midrule
$G^{300}$ & 0.432 & \begin{tabular}[]{@{}c@{}}Cs (0.107)\\ Rb (0.120)\\ K (0.160)\\ Ba (0.168)\\ Sr (0.280)\\ Eu (0.302)\\ Na (0.336)\\ Ca (0.382)\\ Cd (0.403)\\ Hg (0.408)\\ La (0.416)\\ Pr (0.426)\end{tabular} & \begin{tabular}[]{@{}c@{}}Na (0.331)\\ Ag (0.355)\\ Ru (0.363)\\ Os (0.370)\\ Rh (0.376)\\ Zn (0.376)\\ W (0.384)\\ Re (0.397)\\ Mn (0.398)\\ Cd (0.398)\\ Cu (0.400)\\ Hg (0.407)\\ Mg (0.410)\\ Sc (0.415)\\ Mo (0.430)\end{tabular} & 708 \\ \bottomrule
\end{tabular}
\label{table:perovskite_rec}
\end{table}

\section{Kagome Lattice Prototype Structure Recommendations}
\label{kagome_prediction}

The Table~\ref{table:kagome_rec} shows all ion-site recommendations given for A, B, $\text{C}^{\text{1}}$, and $\text{C}^{\text{2}}$ sites by the RS system built from the optimized $G^{100}$, and  with $\mathcal{C} = \text{OQMD}^{\text{current}}$.

\begin{table}[ht]
\centering
\caption{Recommendations for the sites A, B, $\text{C}^{\text{1}}$, and $\text{C}^{\text{2}}$ of the Kagome Lattice \ce{CsV3Sb5} crystal structure prototype, ordered by ion-site distances in the embedding space.}
\begin{tabular}{c c c c c c}
\toprule
 \shortstack{ion-site dist \\threshold} & \shortstack{A site \\(ion-site dist)} & \shortstack{B site \\(ion-site dist)} & \shortstack{$\text{C}^{\text{1}}$ site \\(ion-site dist)} & \shortstack{$\text{C}^{\text{2}}$ site \\(ion-site dist)} & \shortstack{\# of \ce{AB3C5} \\combinations} \\ \midrule
0.328 & \begin{tabular}[]{@{}c@{}}Rb (0.018)\\ K (0.022)\\ Cs (0.027)\\ Tl (0.118)\\ Na (0.130)\\ Ba (0.261)\end{tabular} & \begin{tabular}[]{@{}c@{}}V (0.035)\\ Nb (0.73)\\ Ta (0.085)\\ Mn (0.086)\\ Ti (0.099)\\ Cr (0.103)\\ Mo (0.127)\\ Re (0.139)\\ Fe (0.152)\\ Hf (0.193)\\ W (0.199)\\ Co (0.201)\\ Zr (0.203)\\ Ru (0.203)\\ Os (0.206)\\ Rh (0.209)\\ Ir (0.248)\\ Ni (0.299)\\ Sc (0.299)\\ Li (0.311)\\ Mg (0.312)\\ Hg (0.318)\\ Be (0.322)\end{tabular} & \begin{tabular}[]{@{}c@{}}Sb (0.004)\\ Bi (0.028)\\ As (0.075)\\ P (0.157)\\ Ir (0.241)\\ Pt (0.289)\\ Ta (0.291)\\ Re (0.308)\\ Rh (0.319)\\ Au (0.232)\\ Os (0.325)\end{tabular} & \begin{tabular}[]{@{}c@{}}Sb (0.004)\\ Bi (0.027)\\ As (0.075)\\ P (0.160)\\ Ir (0.229)\\ Pt (0.274)\\ Ta (0.301)\\ Re (0.306)\\ Rh (0.306)\\ Au (0.313)\\ Os (0.318)\\ Pd (0.325)\end{tabular} & 1518 \\ \bottomrule
\end{tabular}
\label{table:kagome_rec}
\end{table}
\bibliography{achemso-demo}